\documentclass[a4paper,12pt]{article}
\usepackage{authblk}
\usepackage[utf8]{inputenc}
\usepackage[margin=0.75in]{geometry}
\usepackage{graphicx}
\usepackage{amsmath}
\usepackage{amsfonts}
\usepackage{subfigure}
\usepackage{caption}

\title{Non-Zero Mean Quantum Wishart Distribution Of Random Quantum States And Application }
\author{Shrobona Bagchi}
\affil{Center for Quantum Information, Korea Institute of Science and Technology, Seoul, 02792, Korea.}
\date{}

\begin{document}
\maketitle

\begin{abstract} 
Random quantum states are useful in various areas of quantum information science. Distributions of random quantum states using Gaussian distributions have been used in various scenarios in quantum information science. One of this is the distribution of random quantum states derived using the Wishart distibution usually used in statistics. This distribution of random quantum states using the Wishart distribution has recently been named as the quantum Wishart distribution \cite{Han}. The quantum Wishart distribution has been found for non-central distribution with a general covariance matrix and zero mean matrix \cite{Han}. Here, we find out the closed form expression for the distribution of random quantum states pertaining to non-central Wishart distribution with any general rank one mean matrix and a general covariance matrix for arbitrary dimensions in both real and complex Hilbert space. We term this as the non-zero mean quantum Wishart distribution. We find out the method for the desired placement of its peak position in the real and complex Hilbert space for arbitrary dimensions. We also show an application of this via a fast and efficient algorithm for the random sampling of quantum states, mainly for qubits where the target distribution is a well behaved arbitrary probability distribution function occurring in the context of quantum state estimation experimental data .
\end{abstract}

\section{Introduction} 
Since the conception of quantum physics, it has undergone many different evolutions and one of this has resulted in the emergence of the modern era of the quantum information science. Many theoretical tools as well as computational tools have been developed to discover new phenomenon as well as use them in various technological applications. In this regard, quantum states constitute one of the most studied aspect in quantum physics, since more knowledge about quantum states enables one to use them as resources in more efficient way in various tasks in quantum information and communication. In this context, the field of random quantum states is particularly useful. For example, they have been shown to be useful in some applications like the numerical testing of the typicality of entanglement in the bipartite quantum states or even in analyzing the efficiency of a noise protection scheme. One of the main theory of random quantum states is based on the Gaussian distribution, namely via the Wishart distribution \cite{collins}. The quantum counterpart of Wishart distribution for random quantum states has been there for the case of zero mean and identity covariance matrix \cite{collins}. Recently it has been extended to the case of the zero mean matrix and a general non-identity covariance matrix \cite{Han}. In this work we further develop this theory by generalizing the quantum Wishart distribution to the case of the non-zero rank one mean matrix and a general covariance matrix for the case of all dimensions and for both real and complex Hilbert space. While we find out the expression for the non-zero mean quantum Wishart distribution, we also develop an algorithm to show its application. In this respect, some algorithms that were already implemented are based on the Markov Chain Monte Carlo (MCMC) and the Hamiltonian Monte Carlo (HMC) techniques \cite{Shang, Seah} and also recently using the zero mean non-central Wishart distribution \cite{Han}. However in the type of the Monte Carlo sampling methods \cite{Shang, Seah}, one typically requires a long time to generate random sample points according to a probability distribution pertaining to arbitrary combination of detector clicks in an experiment. This is because, since we need random samples from quantum state space of an arbitrary landscape, one needs to check the positivity of the generated random point, which is a non-trivial task and checking this for every generated sample point therefore consumes a very long time. Also in MCMC sampling methods, the generated random sample points typically have correlations between them and in some cases, one has to throw away a large number of such correlated sample points \cite{Shang, Seah}. On the other hand, rejection sampling using the uniform distribution in quantum state space offers a viable method for sampling from, however it lacks any kind of flexibility to adapt to any arbitrary landscape and yet show good performance. An efficient algorithm was proposed using the zero mean non-central Wishart distribution using rejection sampling. However, the distribution for that case lacks some flexibility that the non-zero mean non-central quantum Wishart distribution as proposed here has. For example the non-central quantum Wishart distribution with zero mean matrix but non-identity covariance matrix will always have zero density on the boundary of the Bloch ball or the surface of the higher dimensional hypersphere denoting the boundary of the space of quantum state in higher dimension. However, the non-central quantum Wishart distribution with non-zero mean matrix has nonzero density on the boundary, and even sometimes the peak on the boundary. This provides added flexibility. As a result, we explore the option of using the non-zero mean non-central Wishart distribution combined with rejection sampling in the task of random sampling of quantum states according to an arbitrary probability distribution. Other than these, the other existing sampling methods such as those based on induced measure has limited applicability because, often one needs random quantum states according to a very specific target distribution, which is not exactly given by the distributions based on induced measures. As a result, if one needs a random sample from an arbitrary distribution in  quantum state space, one is short of choice for algorithms that take care of all the above issues. Therefore, in this paper, we address the above sampling issues with an algorithm based on rejection sampling and the non-central non-zero mean quantum Wishart distribution. We propose an algorithm that automatically generates physically permissible sample points, does not contain correlations, takes significantly much less time and adapts to arbitrary target distribution from which we wish to sample from. Also apart from the sampling algorithm we believe that more applications can be derived using it for example in the theory of multiple input multiple output quantum channels like the normal non-zero mean Wishart distribution has found use in the multiple input multiple output classical channel theory. As a result, our work provides a way forward to a new direction of research in quantum information science.

The paper is organized as follows. In Section 2, we describe the essential background literature and the necessary definitions. These include the literature on the random matrices based on the non-central Wishart distributions, the maximum likelihood estimator in the quantum state estimation procedure and the sampling technique of acceptance-rejection method and a numerical method for testing the quality of samples. In section 3, we derive our expression for the new distribution for the random quantum states for non-zero rank one mean matrix, which involves a closed form expression and show its different properties for different cases of the state parameter values. In section 4, we show the usefulness of our method by sampling from a number of measurement settings arising in the context of quantum state estimation that includes the trine measurement setting and the tetrahedral measurement setting in a single qubit case and compare with earlier results and in different scenarios. In this section for the purpose of sampling algorithm,  we derive a condition for finding the peak position using two different methods, one using an identity covariance matrix but rank one non-zero mean matrix and another using the non-identity covariance matrix and non-zero mean matrix of rank one. We use the first method of finding peak positions of the proposal probability distributions to show our applications. We support the claim of good quality random samples using numerical technique and evidence. We end with conclusion and discussion on future directions in section 5. 

\section{Background}

\subsection{Wishart random matrices}

There are various ways of generating random density matrices in the field of quantum information theory. One of these methods is based on the distribution of random matrices called the Wishart distribution \cite{Wishart,collins}. In this case, the elements of the random matrices are chosen randomly according to the Gaussian probability distribution with zero mean and unit variance in the usual cases. However, the elements can also be chosen according a non-zero
mean and non-identity covariance matrix. Therefore, we review here some key properties of the Wishart ensemble of random matrices \cite{Wishart,collins} that we will use in our analysis later. The Wishart matrices are random matrices. These matrices are formed according to the rule $W=AA^T$ for the real case and $W=AA^\dagger$  for the complex case, where $W$ is a square matrix of size $n$, whereas the matrix $A$ is in general a rectangular matrix of size $n\times m$ with $m\geq n$. For these random density matrices, one chooses the elements of $A$ as random numbers from the Gaussian probability distribution.
These numbers are usually chosen according to Gaussian distribution with a zero mean, unit variance or with non-zero mean and non-identity covariance matrices. The distribution of these random matrices have been studied in the most general case in the literature \cite{Wishart}. The case where one chooses with non-zero mean is called the non-central Wishart distribution. In this type of distributions, there are two different cases of the Wishart distribution, one where all the entries are real and the other one where all the entries are complex numbers. Now we state the result from the Wishart distribution \cite{Zys,Anderson,Alan} as follows:

\subsubsection{\textbf{The real non-central Wishart distribution}}

The Wishart distribution when the elements of $W$ are all real \cite{Alan}, is given by the following integral
\begin{align}
I(W)=\int_{-\infty}^\infty \delta(W-AA^T)e^{-\frac{1}{2}\mathrm{tr}\{(A^T-M^T)\Sigma^{-1}(A-M)\}} d[AA^T],
\end{align}
where $d[AA^T]=\prod_{ij}W_{ij}$. It has been found that when the elements of the matrix $A$ are all real, then the matrix $W$ has the following distribution \cite{Alan}
\begin{align}
I(W)=\frac{e^{-\frac{1}{2}\mathrm{Tr}(\Sigma^{-1}MM^T)}e^{-\frac{1}{2}\mathrm{Tr}(\Sigma^{-1}W)}}{2^{\frac{dN}{2}}\Gamma_d(\frac{N}{2})|\Sigma|^{\frac{N}{2}}}{\tilde{F}}_1(\frac{N}{2};\frac{1}{4}\Sigma^{-1}MM^T\Sigma^{-1}W)det(W)^{\frac{N-d-1}{2}}
\end{align}
in terms of the Hypergeometric function of the matrix argument, defined as follows
\begin{align}
{\tilde{F}}_1(\frac{N}{2};\mathbf{A})=\sum_{k=0}^\infty\frac{1}{k!}\sum_{\kappa}\frac{C_\kappa(\mathbf{A})}{[\frac{N}{2}]_k}
\end{align}
where, $M$ in called the Mean matrix and $\Sigma$ is the covariance matrix. The elements of $M$ are the values of the mean of the Gaussian distribution from which the element is drawn randomly. Therefore the dimensionality of the matrix $M$ is the same as the dimensionality of the matrix $A$. The covariance matrix $\Sigma$ is the ususal covariance matrix. $\Gamma_d(\frac{N}{2})$ is the generalized gamma function given by $\Gamma_d(\frac{N}{2})=\pi^{\frac{d(d-1)}{4}}\prod_{i=1}^d \Gamma(\frac{N}{2}+1-i)$. $C_\kappa(\mathbf{A})$ is the zonal polynomial which depends through the eigenvalues of the argument matrix $\mathbf{A}$, $\kappa=(k_1,k_2,..,k_n)$, with $k_i$'s being non-negative integers, is a partition of $k$ such that $k_1\geq k_2\geq ..\geq k_n$ and $\sum_{i=1}^n k_i=k$. Also, $[n]_k=\prod_{i=1}^n(n-i+1)_k$, with $(a)_n=a(a+1..(a+n-1))$ denoting the Pochhammer symbol. The matrix $\frac{1}{4}\Sigma^{-1}MM^T\Sigma^{-1}$ is called the non-centrality parameter for the real non-central Wishart distribution since its non zero value determines the deviation from the central Wishart distribution.

\subsubsection{\textbf{The complex non-central Wishart distribution}}
 
The Wishart distribution when the elements of $W$ are complex in general \cite{Alan}, is given by 
\begin{align}
I(W)=\int_{-\infty}^\infty \delta(W-AA^\dagger)e^{-\mathrm{tr}\{(A^\dagger-M^\dagger)\Sigma^{-1}(A-M)\}} d[AA^\dagger],
\end{align}
where $d[AA^\dagger]=\prod_{ij}\mathrm{Re}(W_{ij})\mathrm{Im}(W_{ij})$. 
The distribution when all the entries are complex is given by the following
\begin{align}
I(W)=\frac{e^{-\mathrm{Tr}(\Sigma^{-1}MM^\dagger)}e^{-\mathrm{Tr}(\Sigma^{-1}W)}}{\Gamma_d(N)|\Sigma|^N}{\tilde{F}}_1(N;\Sigma^{-1}MM^\dagger\Sigma^{-1}W)det(W)^{N-d}
\end{align}
where ${\tilde{F}}_1$ is a Hypergeometric function of matrix argument, defined as before in the case of real matrices, only with the arguments of the Hypergeometric function and the Zonal polynomial changed to zonal polynomial of a different argument used here. $\Gamma_d(N)$ is the generalized gamma function given by $\Gamma_d(N)=\pi^{\frac{d(d-1)}{2}}\prod_{i=1}^d \Gamma(N+1-i)$. The matrix $M$ is also the Mean matrix, whose dimensionality is exactly the same as the matrix $A$ whose elements $A_{ij}=AR_{ij}+iAC_{ij}$ are drawn from Gaussian distribution with mean $\mu_{ij}^r+i\mu_{ij}^c$ where the superscripts denote the real and complex parts only. The matrix $\Sigma^{-1}MM^T\Sigma^{-1}$ is called the non-centrality parameter for the case of complex Wishart distribution in the same way as the real Wishart matrix case.

\subsection{Maximum likelihood estimator (MLE) in quantum state estimation and corresponding random sample of quantum states}

For the task of quantum state estimation, in an experiment, an experimenter sends a probe quantum state to be estimated in a measurement set up and performs a measurement whose outcome is registered as a detector click. To analyze the data collected via these detector clicks, we need to know some properties of the measurement set up, which are discussed as follows. A measurement in quantum mechanics is a probability-operator measurement (POM). The POM has outcomes $\Pi_1, \Pi_2, ...,\Pi_k ,$ which are non-negative operators, $\Pi_k \geq 0$, with unit sum, $\sum_k  \Pi_k = 1$. It is known that if $\rho$ is the
actual quantum state, the probability that the $k^{th}$ detector clicks is given by the Born rule, $p_k =\mathrm{tr}(\Pi_k\rho)$. All the possible  
probabilities $p=(p_1,p_2,...,p_k )$ for the chosen POM constitute the total probability space. Given $\{p_k\}$, one can obtain a corresponding density matrix $\rho$. 
If there is a choice among several $\rho$s for the same $p$ (which can
happen when the POM is not informationally complete), we can take one representative, and thus have a one-to- one mapping $\rho $ with $ p$. Because of the one-to-one mapping between states and probabilities, we can therefore identify $p$ with $\rho$. Now we come to the measurement data obtained from detector clicks.
The measurement data D consist of the observed sequence of detector clicks, with $n_k$ clicks of the $k^{th}$ detector after measuring a total number of $N = \sum_k n_ k$ 
copies of the state. The probability of obtaining the measurement data D, if $p$ is the true state, is the point likelihood given by $L(D |p)=p_{n_1}p_{n_2}p_{n_3}p_{n_K}$.
The positivity of $\rho$ and its normalization to unit trace leads to the fact that $p$ satisfies the basic constraints $p_k \geq 0$ and $\sum_k p_k =1$. Since the probabilities result from the POM via the Born rule, the positivity of $\rho$ usually implies further constraints on $p$. Upon multiplying the prior density with the point likelihood for the observed data, we get the (unnormalized) posterior density $wD(p)=w_0(p)L(D|p)$, which we use here as the target distribution for sampling of random quantum states. 
Sampling from the probability space as mentioned above in accordance with the prior is usually relatively simple. However, sampling according to the posterior is generally a difficult task to do. To simplify the usually difficult process of sampling from the arbitrary target distribution arising in the context of posterior distribution corresponding to the measurement data, in this paper, we propose an algorithm which can sample from a posterior density or which we call the target probability density function in a simple and yet a fast way. The quality of samples are also high. The samples obtained for such target probability density function are useful in various areas of quantum information science for example in the numerical integration for obtaining the average of some quantum information theoretic quantity.

\subsection{\textbf{Acceptance-Rejection sampling}}

The acceptance-rejection sampling technique is a procedure to simulate an arbitrary  probability density function $f(x)$. In acceptance-rejection method, the most basic step is to find an alternative probability distribution with density function $g(x)$, 
from which we already have an efficient algorithm for generating from and also such that the function $g(x)$ is `close' to $f(x)$. The algorithm is simple and is described as 
follows. At first, one generates a random variable $Y$ distributed according to probability distribution $g(y)$. Then, one generates a number $U$ randomly which is independent from $Y$ and is drawn randomly from uniform distribution between $0$ and $1$. 
The one checks the condition if $U\leq f(Y)/ cg(Y )$, such that one sets $X = Y$ (accept) or otherwise rejects this point and goes back to first step (reject). These sequence of steps completely describes the acceptance-rejection algorithm. We state here some important points to note of this sampling method. First, since $f(Y )$ and $g(Y )$ are random variables, so is the ratio $f(Y )$ and this ratio is independent of $U$ in $cg(Y)$.
Secondly, the ratio is bounded between $0$ and $1$, i.e., $0< f(Y)/ cg(Y )\leq1$.
Third, the number of times $N$ that the first two steps need to be called (e.g., the number of iterations needed to successfully generate $X$ ) is itself a random variable and has a geometric distribution with success probability $p = P(U \leq f(Y )/ cg(Y )$. Also, along this line we have $P(N = n) = (1-p)n-p, ~ n \geq 1$. Thus on average the number of iterations required is given by $E(N) = 1/p$. In the end we obtain our $X$ as having the conditional distribution of a $Y$ given that the event ${U\leq f(Y)/ cg(Y )}$ occurs. The acceptance-rejection sampling in essence is an algorithm that produces a random sample according to the target distribution which is independent  and identically distributed, unlike some Monte Carlo algorithms which produces correlations between generated sample points.

\subsection{\textbf{Quality of random samples: Bounded likelihood regions}}

The theory of bounded likelihood regions, in the context of random sampling has been discussed in length in \cite{ Shang, Seah, Han}. It will also serve us to determine the quality of our random samples. For our purpose, we discuss in short the main points that are useful for us. A bounded likelihood region in the state space of density matrices is specified by a fraction of the threshold value $L(D|\rho_{ML})$, the fraction being denoted by $\lambda$. The characteristic function of the bounded likelihood region is denoted by the step function as follows:
\begin{align}
\eta_{R_\lambda}(\rho)=\eta(L(D|\rho)-\lambda L(D|\rho_{\rho_{ML}})),
where\\
\eta_{R}(\rho)=1 : \rho\in R, ~~~\eta_{R}(\rho)=0~~~ o.w.
\end{align} 
is called the characteristic function of region $R$ in the quantum state space. 

\subsubsection{\textbf{Size and credibility of bounded likelihood region}}

The size $s_\lambda$ and credibility $c_\lambda$ of a bounded likelihood region $R$ is \begin{align}
s_\lambda=\int_{R_0} d\rho~\eta _{R_\lambda}(\rho),~~~
c_\lambda=\frac{1}{L(D)}\int_{R_0} d\rho~\eta _{R_\lambda}(\rho)L(D|\rho)~~~ |' ~~c_\lambda=\frac{\lambda S_\lambda +\int _{\lambda'=\lambda}^1 S_{\lambda'}d\lambda'}{\int _{\lambda'=0}^1 S_{\lambda'}d\lambda'}.
\end{align}

\subsubsection{\textbf{Numerical program for size and credibility}}

We use the size and credibility of the bounded likelihood regions as defined above to check the quality of samples numerically \cite{Shang, Seah, Han}. For this, we note that the size of the sample can be evaluated by generating a uniform distribution in the state space and evaluating $s_\lambda$ according to $\eta(L(D|\rho)-\lambda L(D|\rho_{ML}))$ and then integrating according to Eq(8) to find the credibility of the sample. This serves as the theoretical credibility to which we compare the credibility of the posterior distribution sample. If these two plots match closely enough, it gives us the evidence that the sample generated according to the posterior distribution is correct. For numerical evaluation, the size is therefore the ratio of states in $R$ to that in the whole state space for a uniform distribution, whereas credibility is calculated by the ratio of states in $R$ to that in the whole state space in the posterior sample.

\section{Results}

In this section, we present our results that range from the derivation of the exact formula or expression for the distributions of random quantum states according rank one non-zero mean matrix and a non-identity general covariance matrix in the both real and complex Hilbert space to the application of the derived distribution function in an efficient sampling of random quantum states for single qubit distribution where the target function is an arbitrary probability distribution function. 

\subsection{Arbitrary distribution in real Hilbert space: Rank-1 Mean Matrix and general Covariance matrix}

We derive an expression for the distribution for rank one Mean matrix and general covariance matrix. For this purpose, we use the properties of the Wishart random matrices found in the statistic literature. We start with the formal expression for the expectation values in the form of integration over the parameter space which will give us the distribution of the random quantum states. For our purpose, we begin with the case of the real Hilbert space. Thus, to calculate the distribution of points in the real Hilbert space, we evaluate the following integration 
\begin{align}
I=E(\delta(\rho-\frac{AA^T}{\mathrm {tr} (AA^T)}))=\int_{-\infty}^\infty \delta(\rho-\frac{AA^T}{\mathrm {tr} (AA^T)})e^{-\frac{1}{2}\mathrm{tr}\{(A^T-M^T)\Sigma^{-1}(A-M)\}} d[AA^T],
\end{align}
where, $d[AA^T]=\prod_{ij}d(A_{ij})$
The above integral can be written as the following
\begin{align}
I=E(\delta(\rho-\frac{AA^T}{\mathrm {tr} (AA^T)}))=\int_{0}^{\infty} \int_{-\infty}^{\infty} \delta(s-\mathrm {tr} (AA^T))\delta(\rho-\frac{AA^T}{s})e^{-\frac{1}{2}\mathrm{tr}\{(A^T-M^T)\Sigma^{-1}(A-M)\}} \mathrm{d}[AA^T]\mathrm{d}s\\
=\int_{0}^{\infty} \int_{-\infty}^{\infty} \delta(s-\mathrm {tr} (AA^T))\delta(\rho-\frac{AA^T}{s})e^{-\frac{1}{2}\mathrm{tr}(\Sigma^{-1}AA^T)}e^{-\frac{1}{2}\mathrm{tr}(\Sigma^{-1}MM^T)}e^{\frac{1}{2}\mathrm{tr}(A^T\Sigma^{-1}M+M^T\Sigma^{-1}A)} \mathrm{d}[AA^T]\mathrm{d}s
\end{align}
We scale $A\rightarrow \sqrt{s}A$. Then the above integral $I$ becomes 
\begin{align}
\int_{0}^{\infty} \int_{-\infty}^{\infty} \delta(s-s\mathrm {tr} (AA^T))\delta(\rho-AA^T)s^{\frac{dN}{2}}e^{-\frac{s}{2}\mathrm{tr}(\Sigma^{-1}AA^T)}e^{-\frac{1}{2}\mathrm{tr}(\Sigma^{-1}MM^T)}e^{\frac{\sqrt{s}}{2}\mathrm{tr}(A^T\Sigma^{-1}M+M^T\Sigma^{-1}A)} \mathrm{d}[AA^T]\mathrm{d}s\\
= \delta(1-\mathrm {tr} (\rho))e^{-\frac{1}{2}\mathrm{tr}(\Sigma^{-1}MM^T)}\int_{0}^{\infty} s^{\frac{dN}{2}-1}e^{-\frac{s}{2}\mathrm{tr}(\Sigma^{-1}\rho)}\int_{-\infty}^{\infty}\delta(\rho-AA^T)e^{\frac{\sqrt{s}}{2}\mathrm{tr}(A^T\Sigma^{-1}M+M^T\Sigma^{-1}A)} \mathrm{d}[AA^T]\mathrm{d}s
\end{align}
Now let us analyze integral of the form below:
\begin{align}
I(W)=E(\delta(W-AA^T))=\int_{-\infty}^{\infty} \delta(W-AA^T)e^{-\frac{1}{2}\mathrm{tr}\{(A^T-M^T)\Sigma^{-1}(A-M)\}} d[AA^T],
\end{align}
We break the above integral in the following form :
\begin{align}
I(W)=\int_{-\infty}^{\infty} \delta(W-AA^T)e^{-\frac{1}{2}\mathrm{tr}(\Sigma^{-1}AA^T)}e^{-\frac{1}{2}\mathrm{tr}(\Sigma^{-1}MM^T)}e^{\frac{1}{2}\mathrm{tr}(A^T\Sigma^{-1}M+M^T\Sigma^{-1}A)} \mathrm{d}[AA^T]\\
=e^{-\frac{1}{2}\mathrm{tr}(\Sigma^{-1}W)}e^{-\frac{1}{2}\mathrm{tr}(\Sigma^{-1}MM^T)}\int_{-\infty}^{\infty} \delta(W-AA^T)e^{\frac{1}{2}\mathrm{tr}(A^T\Sigma^{-1}M+M^T\Sigma^{-1}A)} \mathrm{d}[AA^T]
\end{align}
Now, we use the result of the Wishart density matrices. We know that the integral can be written as an infinite series, as given in the section on background literature. This infinite becomes simpler when the Mean matrix is of rank one. Specifically, for rank-1 Mean matrix, there is only one non-zero eigenvalue of the matrix $MM^T$. Also, we know the following 
\begin{align}
\mathrm{rank}(AB)\leq min\{ \mathrm{rank}(A), \mathrm{rank}(B)\}
\end{align}
Following the above inequality, we see that 
\begin{align}
\mathrm{rank}(\Sigma^{-1}MM^\dagger\Sigma^{-1}W)\leq min\{ \mathrm{rank}(\Sigma^{-1}MM^\dagger), \mathrm{rank}(\Sigma^{-1}W)\}
\end{align}
which, by the same argument implies the following
\begin{align}
\mathrm{rank}(\Sigma^{-1}MM^\dagger\Sigma^{-1}W)\leq  \mathrm{rank}(MM^\dagger)
\end{align}
We know that $\mathrm{rank}(MM^\dagger)=1$. Thus, $\mathrm{rank}(\Sigma^{-1}MM^\dagger\Sigma^{-1}W)=1$. 
Since $\mathrm{rank}(\Sigma^{-1}MM^\dagger\Sigma^{-1}W)=1$, we need to calculate its only nonzero eigenvalue. It is straightforward to calculate the eigenvalue of $\Sigma^{-1}MM^\dagger \Sigma^{-1}W$.
Also, since $\Sigma^{-1}MM^\dagger\Sigma^{-1}W$ contains only one non-zero eigenvalue, therefore $C_\kappa(\Sigma^{-1}MM^\dagger \Sigma^{-1}W)=0$ for all partitions of $k$ having more than one non-zero parts. So only partitions of the form $(k, 0, 0,..0)$ contribute to the summation.  Therefore, for the special case of rank one Mean matrix we have the following:
\begin{align}
{\tilde{F}}_1(\frac{N}{2};\mathbf{\frac{1}{4}\Sigma^{-1}MM^T\Sigma^{-1}W})=\sum_{k=0}^\infty\frac{1}{k!}\frac{C_k(\mathbf{\frac{1}{4}\Sigma^{-1}MM^T \Sigma^{-1}W})}{[\frac{N}{2}]_k}
\end{align}
which is nothing but the following equation
\begin{align}
{\tilde{F}}_1(\frac{N}{2};\mathbf{\frac{1}{4}\Sigma^{-1}MM^T \Sigma^{-1}W})=\sum_{k=0}^\infty\frac{1}{k!}\frac{\xi^{2k}}{[\frac{N}{2}]_k}
\end{align}
where $\xi^2$ is the eigenvalue of the matrix $\frac{1}{4}\Sigma^{-1}MM^T \Sigma^{-1}W$.  Now, we know that $I(W)$ is the Wishart distribution whose exact expression is already given in literature, which is 
\begin{align}
I(W)=\frac{e^{-\frac{1}{2}\mathrm{Tr}(\Sigma^{-1}MM^T)}e^{-\frac{1}{2}\mathrm{Tr}(\Sigma^{-1}W)}}{2^{\frac{dN}{2}}\Gamma_d(\frac{N}{2})|\Sigma|^{\frac{N}{2}}}{\tilde{F}}_1(\frac{N}{2};\frac{1}{4}\Sigma^{-1}MM^T\Sigma^{-1}W)det(W)^{\frac{N-d-1}{2}}
\end{align}
Therefore, comparing the above equation, we get the following
\begin{align}
\int_{-\infty}^{\infty} \delta(W-AA^T)e^{\frac{1}{2}\mathrm{tr}(A^T\Sigma^{-1}M+M^T\Sigma^{-1}A)} \mathrm{d}[AA^T]=\frac{det(W)^{\frac{N-d-1}{2}}}{2^{\frac{dN}{2}}\Gamma_d(\frac{N}{2})|\Sigma|^{\frac{N}{2}}}{\tilde{F}}_1(\frac{N}{2};\frac{1}{4}\Sigma^{-1}MM^T\Sigma^{-1}W)
\end{align}
Now, we use the above expression to replace the value of the integral in the parenthesis in Eq(42) with the modification $M\rightarrow \sqrt{s}M$. Thus, Eq(42) can be evaluated 
\begin{align}
\frac{\delta(1-\mathrm {tr} (\rho))det(\rho)^{\frac{N-d-1}{2}}e^{-\frac{1}{2}\mathrm{tr}(\Sigma^{-1}MM^T)}}{2^{\frac{dN}{2}}\Gamma_d(\frac{N}{2})|\Sigma|^{\frac{N}{2}}}\int_{0}^{\infty} s^{\frac{dN}{2}-1}e^{-\frac{s}{2}\mathrm{tr}(\Sigma^{-1}\rho)}{\tilde{F}}_1(\frac{N}{2};\frac{s}{4}\Sigma^{-1}MM^T\Sigma^{-1}\rho)\mathrm{d}s.
\end{align}
We put $\frac{s}{2}\mathrm{tr}(\Sigma^{-1}\rho)=t$, for which now we have to evaluate the following integral 
\begin{align}
\frac{\delta(1-\mathrm {tr} (\rho))det(\rho)^{\frac{N-d-1}{2}}e^{-\frac{1}{2}\mathrm{tr}(\Sigma^{-1}MM^T)}}{(\mathrm{Tr}(\Sigma^{-1}\rho))^{\frac{dN}{2}}\Gamma_d(\frac{N}{2})|\Sigma|^{\frac{N}{2}}}\int_{0}^{\infty} t^{\frac{dN}{2}-1}e^{-t}{\tilde{F}}_1(\frac{N}{2};\frac{t\Sigma^{-1}MM^T\Sigma^{-1}\rho}{2\mathrm{Tr}(\Sigma^{-1}\rho)})\mathrm{d}t
\end{align}
From now on, we put $C_R= \frac{e^{-\frac{1}{2}\mathrm{tr}(\Sigma^{-1}MM^T)}}{(\mathrm{Tr}(\Sigma^{-1}\rho))^{\frac{dN}{2}}\Gamma_d(\frac{N}{2})|\Sigma|^{\frac{N}{2}}}$. As discussed before, for rank one Mean matrix we can evaluate the above integral by writing down in series as in Eq(52).
Thus using this expression, using Eq(52) and putting the Pochhammer symbol $[\frac{N}{2}]_j=\frac{\Gamma(\frac{N}{2}+j)}{\Gamma(\frac{N}{2})}$ and integrating over $t$, we have 
\begin{align}
I=E(\delta(\rho-\frac{AA^T}{\mathrm {tr} (AA^T)}))=
C_R\delta(1-\mathrm {tr} (\rho))det(\rho)^{\frac{N-d-1}{2}}\int_{0}^{\infty} t^{\frac{dN}{2}-1}e^{-t}\sum_{k=0}^\infty\frac{1}{k!}\frac{(t\xi^2)^{k}}{[\frac{N}{2}]_k}{d}t\\
=C_R\delta(1-\mathrm {tr} (\rho))det(\rho)^{\frac{N-d-1}{2}}\int_{0}^{\infty} t^{\frac{dN}{2}-1}e^{-t}\sum_{k=0}^\infty\frac{1}{k!}\frac{(t\xi^2)^{k}\Gamma(\frac{N}{2})}{\Gamma(\frac{N}{2}+k)}{d}t
\end{align}
where $\xi^2$ is the eigenvalue of the matrix $\frac{\Sigma^{-1}MM^T\Sigma^{-1}\rho}{2\mathrm{Tr}(\Sigma^{-1}\rho)}$.
 
\subsection{\textbf{Distribution in terms of  Whittaker's M function}}

In terms of the Bessel function, we can write  the above integral in more compact form. Before that we note that the form of the modified Bessel function of the first kind is given by 
\begin{align}
I_\nu(z)=(\frac{1}{2}z)^\nu\sum_{k=0}^\infty \frac{(\frac{1}{4}z^2)^k}{k!\Gamma(\nu+k+1)}
\end{align} 
Therefore, using the above form of infinite series in our case, we have the following
\begin{align}
I=C_R\delta(1-\mathrm {tr} (\rho))det(\rho)^{\frac{N-d-1}{2}}\int_{0}^{\infty} t^{\frac{dN}{2}-1}e^{-t}(t\xi^2)^{-\frac{1}{2}(\frac{N}{2}-1)}I_{\frac{N}{2}-1}(2\xi\sqrt{t})\mathrm{d}t\\
=C_R\delta(1-\mathrm {tr} (\rho))det(\rho)^{\frac{N-d-1}{2}}\xi^{-(\frac{N}{2}-1)}\int_{0}^{\infty} t^{\frac{dN}{2}-1-\frac{1}{2}(\frac{N}{2}-1)}e^{-t}I_{\frac{N}{2}-1}(2\xi\sqrt{t})\mathrm{d}t
\end{align}
We know the value of the above integral from literature which is the following
\begin{align}
\int_0^\infty t^{\omega-\frac{1}{2}} e^{\alpha t}I_{2\nu}(2\beta\sqrt{t})dt=\frac{\Gamma(\omega+\nu+\frac{1}{2})\beta^{-1}\alpha^{-\omega}}{\Gamma(2\nu+1)}e^{\frac{\beta^2}{2\alpha}} M_{-\omega,\nu}(\frac{\beta^2}{\alpha}); \mathrm{Real} (\nu)> -1; \mathrm{Real} (\omega+\nu+\frac{1}{2})>0
\end{align}
Therefore, here we have $\alpha=-1$, $\beta=\xi$, $\nu=\frac{N}{4}-\frac{1}{2}$, $\omega=\frac{N}{2}(d-\frac{1}{2})$. It is straightforward to see that the required constraints are satisfied by these constants, (since we take $N\geq 3$ for non-singular real Wishart distribution) and thus we obtain the following distribution
\begin{align}
f_W(\rho)=(-1)^{-\frac{N}{2}(d-\frac{1}{2})} C_R\delta(1-\mathrm {tr} (\rho))det(\rho)^{\frac{N-d-1}{2}}\frac{\Gamma(\frac{dN}{2})\xi^{-\frac{N}{2}}}{\Gamma(\frac{N}{2})}e^{\frac{-\xi^2}{2}} M_{-\frac{N}{2}(d-\frac{1}{2}),\frac{N}{4}-\frac{1}{2}}(-\xi^2)
\end{align}

\subsection{\textbf{Distribution in terms of exponential and finite polynomial}}

For the cases when $N$ is odd for $d$ even and for all $N$ when $d$ is odd, we begin from 
\begin{align}
I=C_R\delta(1-\mathrm {tr} (\rho))det(\rho)^{\frac{N-d-1}{2}}\xi^{-(\frac{N}{2}-1)}\int_{0}^{\infty} t^{\frac{dN}{2}-1-\frac{1}{2}(\frac{N}{2}-1)}e^{-t}I_{\frac{N}{2}-1}(2\xi\sqrt{t})\mathrm{d}t
\end{align}
After this we make the substitution $\sqrt{t}=t_1$. Then we have the following
\begin{align}
I=2C_R\delta(1-\mathrm {tr} (\rho))det(\rho)^{\frac{N-d-1}{2}}\xi^{-(\frac{N}{2}-1)}\int_{0}^{\infty} {t_1}^{dN-\frac{N}{2}}e^{-{t_1}^2}I_{\frac{N}{2}-1}(2\xi t_1)\mathrm{d}t_1
\end{align}
For $N$ even, we can also express ${t_1}^{N(d-1)}e^{-p{t_1}^2}=(-1)^{\frac{N}{2}(d-1)}(\frac{\mathrm{d}}{\mathrm{d}{p}})^{\frac{N}{2}(d-1)}e^{-p{t_1}^2}$. This can also be done for any value of $N$ when $d$ is odd. Putting this value, we have 
\begin{align}
I=2(-1)^{\frac{N}{2}(d-1)}C_R\delta(1-\mathrm {tr} (\rho))det(\rho)^{\frac{N-d-1}{2}}\xi^{-(\frac{N}{2}-1)}\displaystyle{\lim_{p \to 1}}
\Big(\frac{\mathrm{d}}{\mathrm{d}{p}}\Big)^{\frac{N}{2}(d-1)}\int_{0}^{\infty}{t_1}^{\frac{N}{2}}e^{-p{t_1}^2}I_{\frac{N}{2}-1}(2\xi t_1)\mathrm{d}t_1
\end{align}
We know the value of the above integral from literature which is the following
\begin{align}
\int_0^\infty q^{\nu+1}I_\nu(cq)e^{-pq^2}dq=\frac{c^\nu}{(2p)^{\nu+1}}e^{\frac{c^2}{4p}}; \mathrm{Real} (\nu)> -1; \mathrm{Real} (p)>0
\end{align}
 which we put and also put $\mathrm{tr}(\rho)=1$and get the closed form answer as follows 
\begin{align}
I=C_R (-1)^{\frac{N}{2}(d-1)}det(\rho)^{\frac{N-d-1}{2}}\displaystyle{\lim_{p \to 1}}
\Big(\frac{\mathrm{d}}{\mathrm{d}{p}}\Big)^{\frac{N}{2}(d-1)}\Big[p^{-\frac{N}{2}}e^{\frac{\xi^2}{p}}\Big]
\end{align}

\subsection{Distribution in complex space: } In the previous sections, we derived the expression for random quantum states for the case of real Hilbert space. Here, we derive the expression for complex Hilbert spae, using almost similar technique with also a few modifications to suit the case of complex Hilbert space. For this purpose, we first state the notations we use for our derivations. 

\subsubsection{Rank-1 Mean matrix and Non-Identity covariance}
The distribution of random quantum states in the quantum state space when they have complex entries according to Gaussian distribution with rank one mean matrix 
and a non-identity covariance matrix. The proof of this is very similar to that for the real entries, only differing in some factors and conditions. However, 
for the sake of completeness we give the proof here.
We state here the notations we have used in our derivations of the distributions of random density matrices in the complex Hilbert space. 
We choose the elements according to Gaussian distribution. They are given as follows.
$A$ is a $d\times N$ matrix, where $N$ denotes the number of columns and $d$ is the dimension of the density matrix.
$M$ is the matrix with its elements $M_{ij}=MR_{ij}+iMI_{ij}$ representing the mean of the Gaussian distribution from which the element $A_{ij}$ is 
drawn where $MR_{ij}$ is the mean of the real part of $A_{ij}$ and $MI_{ij}$ is the mean of the imaginary part of $A_{ij}$ . The dimension of $M$ is same as that of $A$.
$\Sigma$ is the covariance matrix of elements of $A$.
The matrix $\Sigma^{-1}MM^T\Sigma$ is called the non-centrality parameter, since it denotes the deviation of the distribution away from the origin.

\subsection{\textbf{Derivation of the distribution in complex Hilbert space}}

Let us note that the distribution in the complex space can be reformulated in terms of the integration of complex random matrices, since the complex matrices are isomorphic to real 
matrices of double dimensions. Let the Mean matrix be $M$ which has elements of the form $r_{h,k}e^{i\theta}$ and let the matrix be of rank 1, which means it has linearly dependent 
rows. We also take the covariance matrix to be $\Sigma$. Now we derive the distribution of random quantum states in the whole complex Hilbert space for any dimension, number of 
columns, when the elements of $A$ are chosen in i.i.d. manner according to Gaussian distribution with mean $\mu_{i,j}$ and variance $\sigma_{i,j}$ with the constraint that the Mean matrix is of rank-1.
\begin{align}
\rho=\frac {AA^\dagger}{\mathrm{Tr}(AA^\dagger)}
\end{align}
We will use the above expression to derive our distribution for the random quantum states.
For this, we note that the distribution of the quantum states is given by 
\begin{align}
I=E(\delta(\rho-\frac{AA^\dagger}{\mathrm {tr} (AA^\dagger)}))=\int \delta(\rho-\frac{AA^\dagger}{\mathrm {tr} (AA^\dagger)})e^{-\mathrm{tr}\{(A^\dagger -M^\dagger)\Sigma^{-1}(A-M)\}}d[AA^\dagger],
\end{align}
where, $d[AA^\dagger]=\prod_{ij}d(A_{ij})d(A_{ij})^*$
The above integral can be written as the following:
\begin{align}
I
=\int_{0}^{\infty} \int_{-\infty}^{\infty} \delta(s-\mathrm {tr} (AA^\dagger))\delta(\rho-\frac{AA^\dagger}{s})e^{-\mathrm{tr}(\Sigma^{-1}A A^\dagger)}e^{-\mathrm{tr}(\Sigma^{-1}M M^\dagger)}e^{\mathrm{tr}(A^\dagger \Sigma^{-1}M+M^\dagger\Sigma^{-1}A)} \mathrm{d}[AA^\dagger]\mathrm{d}s
\end{align}
We scale $A\rightarrow \sqrt{s}A$. Then the above integral becomes 
\begin{align}
I=\delta(1-\mathrm {tr} (\rho))e^{-\mathrm{tr}(\Sigma^{-1}MM^\dagger)}\int_{0}^{\infty} s^{dN-1}e^{-s\mathrm{tr}(\Sigma^{-1}\rho)}\int_{-\infty}^{\infty}\delta(\rho-AA^\dagger)e^{-\sqrt{s}\mathrm{tr}(A^\dagger\Sigma^{-1}M+M^\dagger\Sigma^{-1}A)} \mathrm{d}[AA^\dagger]\mathrm{d}s
\end{align}
Following the same arguments as before about the series for rank-1 Mean matrix, we obtain 
\begin{align}
I(W)
=\frac{det(W)^{N-d}e^{-\mathrm{Tr}(\Sigma^{-1}MM^\dagger)}e^{-\mathrm{Tr}(\Sigma^{-1}W)}}{\Gamma_d(N)|\Sigma|^N}\sum_{k=0}^\infty\frac{\mu^{2k}}{k!(N)_k}
\end{align}
Now, we have $(N)_k$ the Pocchammer symbol defined as $(N)_k=\frac{\Gamma(N)}{\Gamma(N+k)}$, and $\mu^2$ is the eigenvalue of $\mathbf{\Sigma^{-1}MM^\dagger \Sigma^{-1}W}$. 
Putting this value, we obtain the following distribution
\begin{align}
I(W)
=\frac{\Gamma(N)det(W)^{N-d}e^{-\mathrm{Tr}(\Sigma^{-1}MM^\dagger)}e^{-\mathrm{Tr}(\Sigma^{-1}W)}}{\Gamma_d(N)|\Sigma|^N}\sum_{k=0}^\infty\frac{\mu^{2k}}{k!\Gamma(N+k)}
\end{align}
We follow the similar algebraic steps as in the case of real Wishart matrices followed by scaling $M\rightarrow \sqrt{s}M$, then taking $s\mathrm{tr}(\Sigma^{-1}\rho)=t$, we obtain
\begin{align}
I
=\frac{\Gamma(N)\delta(1-\mathrm {tr} (\rho))e^{-\mathrm{tr}(\Sigma^{-1}MM^\dagger)}det(\rho)^{N-d}}{(\mathrm{tr}(\Sigma^{-1}\rho)^{dN})\Gamma_d(N)|\Sigma|^N}\int_{0}^{\infty} t^{dN-1}e^{-t}\sum_{k=0}^\infty\frac{(\sqrt{t}\xi)^{2k}}{k!\Gamma(N+K)}\mathrm{d}[AA^\dagger]\mathrm{d}t
\end{align}
where $\xi^2$ is the eigenvalue of the matrix $\frac{\Sigma^{-1}MM^\dagger\Sigma^{-1}\rho}{\mathrm{tr}(\Sigma^{-1}\rho)}$.
Now we make use of the modified Bessel function of first kind.
\begin{align}
I= \frac{\Gamma(N)\delta(1-\mathrm {tr} (\rho))e^{-\mathrm{tr}(\Sigma^{-1}MM^\dagger)}det(\rho)^{N-d}}{(\mathrm{tr}(\Sigma^{-1}\rho)^{dN})\Gamma_d(N)|\Sigma|^N}\int_{0}^{\infty} t^{dN-1-(\frac{N-1}{2})}e^{-t}\xi^{-(N-1)}I_{N-1}(2\sqrt{t}\xi)\mathrm{d}[AA^\dagger]\mathrm{d}t
\end{align}
We make the transformation $t=t_1^2$, then having to evaluate the following
\begin{align}
I= \frac{2\xi^{-(N-1)}\Gamma(N)\delta(1-\mathrm {tr} (\rho))e^{-\mathrm{tr}(\Sigma^{-1}MM^\dagger)}det(\rho)^{N-d}}{(\mathrm{tr}(\Sigma^{-1}\rho)^{dN})\Gamma_d(N)|\Sigma|^N}\int_{0}^{\infty} t_1^{N(2d-1)}e^{-t_1^2}I_{N-1}(2t_1\xi)\mathrm{d}[AA^\dagger]\mathrm{d}t_1
\end{align}

\subsubsection{\textbf{Distribution in terms of infinite series}}

Using the above expression, and integrating over $s$, we obtain the final distribution of density matrices as :
\begin{align}
E(\delta(\rho-\frac{AA^\dagger}{\mathrm{tr}(AA^\dagger)}))
=\frac{\Gamma(N)det(\rho)^{N-d}e^{-\mathrm{Tr}(\Sigma^{-1}MM^\dagger)}e^{-\mathrm{Tr}(\Sigma^{-1}\rho)}}{\Gamma_d(N)|\Sigma|^N}\sum_{k=0}^\infty\frac{\xi^{2k}\Gamma(dN+k)}{k!\Gamma(N+k)}
\end{align}

\subsubsection{\textbf{Distribution in terms of Whittaker's M function}}

As before, here we have $\alpha=-1$, $\beta=\xi$, $\nu=\frac{N}{2}-\frac{1}{2}$, $\omega=N(d-\frac{1}{2})$ which satisfy necessary constraints. Thus we obtain the following distribution
\begin{align}
E(\delta(\rho-\frac{AA^\dagger}{\mathrm{tr}(AA^\dagger)}))=C_C\frac{\Gamma(Nd)\xi^{-1}(-1)^{-N(d-\frac{1}{2})}}{\Gamma(N)}e^{\frac{-\xi^2}{2}} M_{-N(d-\frac{1}{2}),\frac{N}{2}-\frac{1}{2}}(-\xi^2)
\end{align}

\subsubsection{\textbf{Distribution in terms of exponential and finite polynomial}}

In the complex case, we can always write it in this form irrespective of even or odd value of the number of columns. But, at first we put the constant 
\begin{align}
C_C= \frac{\Gamma(N)e^{-\mathrm{tr}(\Sigma^{-1}MM^\dagger)}}{(\mathrm{tr}(\Sigma^{-1}\rho)^{dN})\Gamma_d(N)|\Sigma|^N}
\end{align}
We see that as before, the expression in the integral is in a standard form whose value exists in the literature which we use and obtain the expression in the following form
\begin{align}
I=C_C(-1)^{N(d-1)}\delta(1-\mathrm {tr} (\rho))\mathrm{det}(\rho)^{N-d}\displaystyle{\lim_{p \to 1}}(\frac{\mathrm{d}}{\mathrm{d}{p}})^{N(d-1)}[p^{-N}e^{\frac{\xi^2}{p}}]
\end{align}

\subsection{\textbf{Example: The case for qubits}}

In this section, we demonstrate the more explicit form of the derived distribution for the case of qubits using the parameters of the Bloch vector used to parameterize a quantum state. We show this for the case of both the real and complex Bloch vector parameters, and for different values of the mean values but an identity covariance matrix for simplicity. The examples for the non-identity covariance can be simply found out using the similar method as presented here.

\subsubsection{\textbf{Real Bloch plane}}

We have $\xi^2$ as the eigenvalue of the matrix $\frac{\Sigma^{-1}MM^T\Sigma^{-1}\rho}{2\mathrm{Tr}(\Sigma^{-1}\rho)}$. 
For all practical purposes of using a non-zero mean for sampling, we can take the covariance matrix to be identity. For number of columns $N$, the rank one Mean Matrix has the 
following form 
\begin{align}
M=
\begin{bmatrix}
\mu && \mu&& \mu&& \mu && ...\\
\mu&& \mu&& \mu&&\mu && ...
\end{bmatrix}_{2\times N}
\end{align}
Then, we have $\xi^2$ as the eigenvalue of $\frac{MM^T\rho}{2}$, which is $\frac{N\mu^2}{2}(1+x)$.
Putting this in Eq(62) we obtain the distribution as 
\begin{align}
C_R (-1)^{\frac{N}{2}}det(\rho)^{\frac{N-3}{2}}\displaystyle{\lim_{p \to 1}}
\Big(\frac{\mathrm{d}}{\mathrm{d}{p}}\Big)^{\frac{N}{2}}\Big[p^{-\frac{N}{2}}e^{\frac{N\mu^2(1+x)}{2p}}\Big]
\end{align}

\subsubsection{\textbf{Complex Bloch Ball}}
Here, we obtain the expression for the distribution of qubits on the real unit disc of the Bloch Ball, 
with an identity covariance matrix for some specific values of the number of columns of the matrix $A$. For $N=4$ and taking the following
mean matrix
\begin{align}
M=\sqrt{2}e^{\frac{i\pi}{4}}\begin{bmatrix}
\mu & \mu&\mu & \mu\\
\mu & \mu&\mu & \mu
\end{bmatrix}
\end{align}
we have $\xi^2$ the eigenvalue of the matrix $\frac{\Sigma^{-1}MM^\dagger\Sigma^{-1}\rho}{\mathrm{Tr}(\Sigma^{-1}\rho)}$ which is
$\xi^2=2N\mu^2(1+x)$. Using this we obtain the distribution as 
\begin{align}
f_W(\rho)
=C_C(\mathrm{det}(\rho))^{N-2}\displaystyle{\lim_{p \to 1}}(\frac{\mathrm{d}}{\mathrm{d}{p}})^{N}[p^{-N}e^{\frac{2N\mu^2(1+x)}{p}}]
\end{align}
\begin{figure}
    \centering
    \includegraphics[scale=0.75]{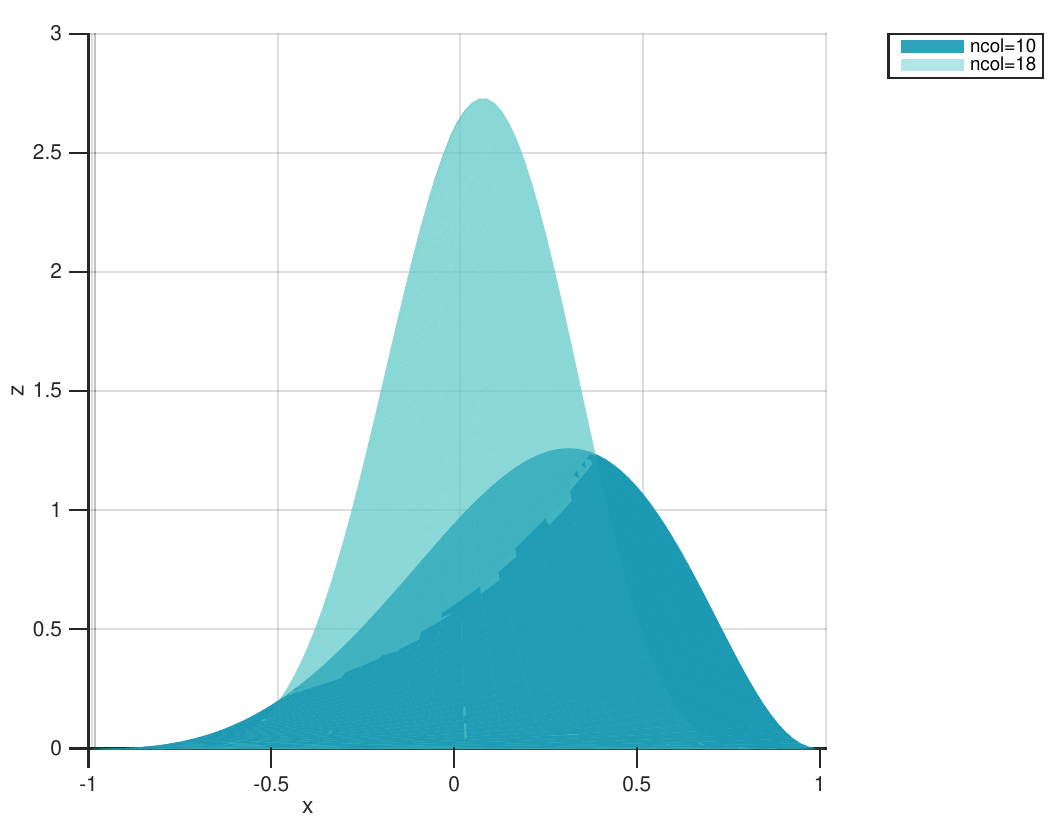}
    \caption{Plot of the cross-section of the probability distribution of random quantum states according to non-zero mean quantum Wishart distribution in two Bloch plane with number of columns 18 and 10 but for the same mean value $\mu=1$. }
    \label{fig:enter-label}
\end{figure}

\subsubsection{\textbf{Non-zero density on the boundary}}

Unlike the above case, if $ncol=3$ and $ncol=2$ for real and complex Wishart distributions respectively and the $\mu$ is nonzero, then clearly the density is non-zero on the boundary, and as one increases or decreases the value of $\mu$, the height of the peak of the distribution changes accordingly, as evident from the complex Wishart distributed density of random quantum states given as follows
\begin{align}
f_W(\rho)=\frac{e^{4\mu^2(x-1)(3+4\mu^2(1+x)(3+2\mu^2(1+x)))}}{4\pi}
\end{align}
It is easy to see that the peak also lies exactly on the $x$ axis and on the boundary,i.e., at $x=1$. Similar result holds for real Wishart distributed random density matrices.
\begin{figure}
    \centering
    \includegraphics[scale=0.75]{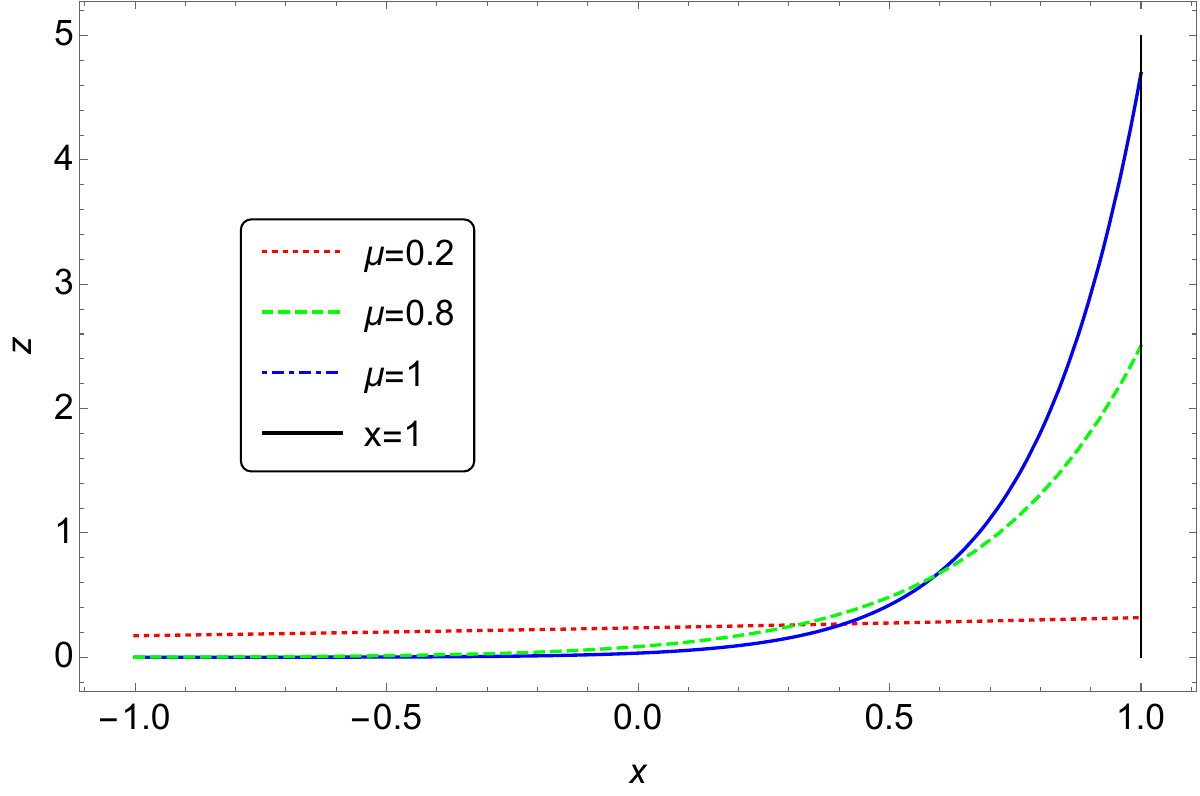}
    \caption{Plot of the cross-section of the probability distribution of random quantum states according to non-zero mean quantum Wishart distrbution in two Bloch plane, for different values of the mean $\mu$ denoted by the legends in the plot, such that one has a non-zero density on the boundary of the Bloch plane here.}
    \label{fig:enter-label}
\end{figure}

\subsection{\textbf{Method to fix desired peak position}}

Now, since we have derived the distribution function of the random density matrices according to the non-central Wishart distribution, let us now analyze how we can fix the peak position at a desired location. In our case, we need to handle only the $N$ and $\mu$ as we take the covariance matrix to be identity. We now outline the following simple steps to do the above.
At first, find the peak $\{r, \theta_1,  ....\theta_n\}$ of the target distribution in the state space .
Then, choose an appropriate $N$ that approximately gives a height close to that of the target distribution peak. After this, form the following equation
{\small\begin{align}
\frac{d}{dq}f(\rho, N, d, \mu)|_{q=r}=0\nonumber
\end{align}}
In the above equation now the only variable is $\mu$, since we have already put the values of $q$, $d$ and $N$ after the differentiation.
Then, numerically find the value of $\mu$ for which the equation holds. It is very easy to see that the value of $\mu$ can be chosen at arbitrarily high precision such that it 
obeys the above equation. This process just takes about a second to solve. After one finds the appropriate value of $\mu$, we put it back in the distribution function of the random density matrix. Evidently, this proposal distribution now has a peak at $q=r$.
After this, perform unitary operation or rotate the sample points accordingly to match the rest of the peak position. Use of unitary transformation is more suitable, without going to Bloch vector space. This gives us the exact analytical form of the proposal distribution.
The above method is a generic method for any value of $N$ and $d$, and thus can be performed in all dimensions without any increase in complexity of the calculation of the solution.
We outline the method with an example here. Take for example $\{r=0.5, \theta=\frac{\pi}{4}\}$ to be position of the peak on the real Bloch plane and choose $N=4$. We have the equation $\frac{d}{dx}f(\rho, 4, 2, \mu)|_{x=0.5}=0\nonumber$.
We know $f(\rho, N, d, \mu)=\frac{\Gamma(2)(1-x^2-z^2)^{\frac{1}{2}}e^{2\mu^2x}}{\Gamma_d(2)}\Big[2\mu^4(1+x)^2+6\mu^2(1+x)+3\Big]$. Therefore, we have the following 
{\small\begin{align}
\frac{d}{dx}\Big [(1-x^2)^{\frac{1}{2}}e^{2\mu^2x}\big[2\mu^4(1+x)^2+6\mu^2(1+x)+3\big]\Big ]_{x=0.5}=0\nonumber
\end{align}}
Numerically we find the value of $\mu_{peak}=0.438946$ for which the equation holds. Rotate the sample points accordingly to match the rest of the peak position. Alternatively one can perform a unitary transformation. The final form of 
the proposal distribution is thus the following 
{\small\begin{align}
\frac{\Gamma(2)(1-r^2)^{\frac{1}{2}}e^{2\mu_{peak}^2(1+rcos(\theta-\frac{\pi}{4}))}}{\Gamma_d(2)}\Big[2\mu_{peak}^4(1+rcos(\theta-\frac{\pi}{4}))^2+6\mu_{peak}^2(1+rcos(\theta-\frac{\pi}{4}))+3\Big]
\end{align}}
Also apart from the above method, which is particularly suited for the qubit space, we outline another method via the calculus of variations and the usage of the covariance matrix and the mean matrix which lets us fix a desired peak position in the Wishart distributions according to a target distribution. The method is outlined as follows.
 
\subsubsection{\textbf {Condition for the stationary point}}

For finding $\Sigma, M$ matrices such that we get a stationary point for $\rho=\rho_{peak}$ , the following equation has to be satisfied. We have taken the log of the function, therefore for infinitesimal change $\delta\rho$ in the quantum state space, we obtain the following condition for stationary point 

\begin{align}
\delta \Big[\log[ \frac{\Gamma(N)e^{-\mathrm{tr}(\Sigma^{-1}MM^\dagger)}}{(\mathrm{tr}(\Sigma^{-1}\rho)^{dN})\Gamma_d(N)|\Sigma|^N}\mathrm{det}(\rho)^{N-d}\displaystyle{\lim_{p \to 1}}(\frac{\mathrm{d}}{\mathrm{d}{p}})^{N(d-1)}[p^{-N}e^{\frac{\xi^2}{p}}] \Big ]=0
\end{align}
Removing the constants which will not be essential for finding the stationary point we obtain the following 

\begin{align}
\delta \Big[\log[ \frac{\mathrm{det}(\Sigma^{-1}\rho)^{N-d}}{(\mathrm{tr}(\Sigma^{-1}\rho)^{dN})}\displaystyle{\lim_{p \to 1}}(\frac{\mathrm{d}}{\mathrm{d}{p}})^{N(d-1)}[p^{-N}e^{\frac{\xi^2}{p}}] \Big ]=0
\end{align}
Breaking the above equation in the following form
\small{\begin{align}
\delta \Big[\log[ \frac{(\mathrm{det}(\Sigma^{-1}\rho))^{N-d}}{(\mathrm{tr}(\Sigma^{-1}\rho)^{dN})}]\Big]+\delta \Big[\log[\displaystyle{\lim_{p \to 1}}(\frac{\mathrm{d}}{\mathrm{d}{p}})^{N(d-1)}[p^{-N}e^{\frac{\xi^2}{p}}]] \Big ]=0\\ \nonumber 
\Rightarrow (N-d)\delta \Big [\log[ (\Sigma^{-1}\mathrm{det}(\rho)) ]\Big] -dN\delta\Big[ [\log(\mathrm{tr}(\Sigma^{-1}\rho))]\Big]+\delta \Big[\log[\displaystyle{\lim_{p \to 1}}(\frac{\mathrm{d}}{\mathrm{d}{p}})^{N(d-1)}[p^{-N}e^{\frac{\xi^2}{p}}] ]\Big ]=0\\ \nonumber 
\Rightarrow \mathrm{Tr}\Big [ \{(N-d)\rho^{-1}-dN\frac{\Sigma^{-1}}{\mathrm{tr}(\Sigma^{-1}\rho)}+T_1(\frac{\Sigma^{-1}MM^\dagger \Sigma^{-1}}{\mathrm{Tr}(\Sigma^{-1}\rho)}- \frac{\mathrm{Tr}(\Sigma^{-1}MM^\dagger \Sigma^{-1}\rho)\Sigma^{-1}}{\mathrm{Tr}(\Sigma^{-1}\rho)^2})\}\delta\rho\Big ]=0\nonumber
\end{align}}
where we have $T_1=\frac{\sum_j C'_j \xi^{j}}{\sum_k C_k \xi^{2k}}$, where $\xi^2=\frac{\mathrm{Tr}(\Sigma^{-1}MM^\dagger \Sigma^{-1}\rho)}{\mathrm{Tr}(\Sigma^{-1}\rho)})$ and $\sum_k C_k \xi^{2k}=\displaystyle{\lim_{p \to 1}}(\frac{\mathrm{d}}{\mathrm{d}{p}})^{N(d-1)}[p^{-N}e^{\frac{\xi^2}{p}}]$. For this to be zero for arbitrary $\delta\rho$ at $\rho=\rho_p$, we have the following 
\small{\begin{align}
 (N-d)\rho^{-1}_p-dN\frac{\Sigma^{-1}}{\mathrm{tr}(\Sigma^{-1}\rho_p)}+T_{1p}(\frac{\Sigma^{-1}MM^\dagger \Sigma^{-1}}{\mathrm{Tr}(\Sigma^{-1}\rho_p)}- \frac{\mathrm{Tr}(\Sigma^{-1}MM^\dagger \Sigma^{-1}\rho_p)\Sigma^{-1}}{\mathrm{Tr}(\Sigma^{-1}\rho_p)^2}) =\mbox{m}I\nonumber
\end{align}}
where $T_{1p}=T_1$ at $\rho=\rho_p$. Multiplying by $\rho_p$ on both sides, we have 
\small{\begin{align}
 (N-d)I-dN\frac{\Sigma^{-1}\rho_p}{\mathrm{tr}(\Sigma^{-1}\rho_p)}+T_{1p}(\frac{\Sigma^{-1}MM^\dagger \Sigma^{-1}\rho_p}{\mathrm{Tr}(\Sigma^{-1}\rho_p)}- \frac{\mathrm{Tr}(\Sigma^{-1}MM^\dagger \Sigma^{-1}\rho_p)\Sigma^{-1}\rho_p}{\mathrm{Tr}(\Sigma^{-1}\rho_p)^2}) =\mbox{m}\rho_p\nonumber
\end{align}}
Now we take the trace on both sides to have 
\small{\begin{align}
 (N-d)d-dN =\mbox{m}\nonumber
 \Rightarrow m=-d^2\nonumber
\end{align}}
We put this in the equation and get 
\small{\begin{align}
 (N-d)\rho^{-1}_p-dN\frac{\Sigma^{-1}}{\mathrm{tr}(\Sigma^{-1}\rho_p)}+T_{1p}(\frac{\Sigma^{-1}MM^\dagger \Sigma^{-1}}{\mathrm{Tr}(\Sigma^{-1}\rho_p)}- \frac{\mathrm{Tr}(\Sigma^{-1}MM^\dagger \Sigma^{-1}\rho_p)\Sigma^{-1}}{\mathrm{Tr}(\Sigma^{-1}\rho_p)^2}) =-d^2I\nonumber
\end{align}}
If we make the transformation $\Sigma^{-1}\rightarrow \frac{\Sigma^{-1}}{\mathrm{Tr}(\Sigma^{-1}\rho_p)}=\Sigma_1^{-1} $, then the density function is changed, so we have to modify $M\rightarrow \sqrt{\mathrm{Tr}(\Sigma^{-1}\rho_p)}M=M_1$ such that the density function is unchanged. This does not violate the only constraint of rank one from the mean matrix.  In this case the above equation is modified into the following 
\small{\begin{align}
 (N-d)\rho^{-1}_p-dN\Sigma_1^{-1}+T_{1p}(\frac{\Sigma_1^{-1}M_1M_1^\dagger \Sigma_1^{-1}}{\mathrm{Tr}(\Sigma_1^{-1}\rho_p)}- \frac{\mathrm{Tr}(\Sigma_1^{-1}M_1M_1^\dagger \Sigma_1^{-1}\rho_p)\Sigma_1^{-1}}{\mathrm{Tr}(\Sigma_1^{-1}\rho_p)}) =-d^2I\nonumber
\end{align}}
such that the condition of the maximum is unchanged. This also shows that we can have many combinations of the mean matrix and the covariance matrix that will give us the same peak position in the state space. We can try to find one such combination. Now let us choose a mean matrix independently of the covariance matrix. Also in the above equation the let us choose the term $\frac{\Sigma_1^{-1}M_1M_1^\dagger \Sigma_1^{-1}}{\mathrm{Tr}(\Sigma_1^{-1}\rho_p)}=M_2M_2^\dagger$ where $M_2$ is a rank one matrix, since the rank of the product of the matrices on the left hand side is also of rank one. This can always be chosen since the matrix $\Sigma_1^{-1}$ is Hermitian and positive semidefinite as the properties of the covariance matrix. The matrix $M_2$ now acts as a constant which we can choose. Also, note that any all the choices of rank one $M2$ will exhaust of all the possible solutions of the peak position, since $\Sigma_1M_1$ is always of rank one. Therefore putting this in the above equation we get the following
\small{\begin{align}
 (N-d)\rho^{-1}_p-dN\Sigma_1^{-1}+T_{1p}(M_2M_2^\dagger- \mathrm{Tr}(M_2M_2^\dagger\rho_p)\Sigma_1^{-1}) =-d^2I\nonumber\\
 \Rightarrow  (N-d)\rho^{-1}_p+d^2I+T_{1p}(M_2M_2^\dagger ) =dN\Sigma_1^{-1}+T_{1p}\mathrm{Tr}(M_2M_2^\dagger\rho_p)\Sigma_1^{-1}\nonumber\\
 \Rightarrow \Sigma_1=\frac{dN+T_{1p}\mathrm{Tr}(M_2M_2^\dagger\rho_p)}{(N-d)\rho^{-1}_p+d^2I+T_{1p}(M_2M_2^\dagger )}\Rightarrow \Sigma_1=\frac{Nd}{N-d}\Big[\frac{1+T_{2p}\mathrm{Tr}(M_2M_2^\dagger\rho_p)}{\rho^{-1}_p+\frac{d^2}{N-d}I+T_{2p}(M_2M_2^\dagger )}\Big] \nonumber\\
 \Sigma_1=\frac{Nd}{N-d}\Big[\frac{1+T_{2p}\mathrm{Tr}(M_2M_2^\dagger\rho_p)}{\rho^{-1}_p+\frac{d^2}{N-d}+T_{2p}(M_2M_2^\dagger )}\Big]\nonumber
\end{align}}
where $T_{2p}=\frac{T_{1p}}{Nd}$ . It is easy to see that the $\Sigma_1$ obtained this way is Hermitian and positive semidefinite as long as $\rho_p$ is of full rank and is invertible. After finding $\Sigma_1$, we can put it in the following equation to find $M_1$ as follows 
\small{\begin{align}
\frac{\Sigma_1^{-1}M_1M_1^\dagger \Sigma_1^{-1}}{\mathrm{Tr}(\Sigma_1^{-1}\rho_p)}=M_2M_2^\dagger\Rightarrow M_1M_1^\dagger=\mathrm{Tr}(\Sigma_1^{-1}\rho_p)
 \Sigma_1M_2M_2^\dagger \Sigma_1\Rightarrow M_1=\sqrt{\mathrm{Tr}(\Sigma_1^{-1}\rho_p)} \Sigma_1M_2
\end{align}}

Also by construction we have $\mathrm{Tr}(\Sigma_1^{-1}\rho_p)=1$, so our solution becomes $M_1= \Sigma_1M_2$. We can do the above by using the property that the covariance matrix is Hermitian and positive semidefinite. It is clear that the above combination of $\{\Sigma_1, M_1\}$ satisfies he condition of stationary point without violating any constraint. Also it can be seen that when we take $M_1=M_2=0$, i.e. the mean matrix is a null matrix, we get back the peak position as given by the covariance matrix only as before as the following 
\begin{align}
\Sigma_1=\frac{Nd}{N-d}\Big[\frac{1}{\rho^{-1}_p+\frac{d^2}{N-d}}\Big] \nonumber
\end{align}
This coincides with the result which has been used in the companion paper. 

\section{\textbf{Results on Sampling}}

In this section, we provide some numerical examples to show the applicability of the mathematical expression for the non-zero mean quantum Wishart distribution in a sampling algorithm for random quantum states according to an arbitrary probability distribution satisfying some desirable characteristics. The algorithm is showcased by different examples as follows.

\subsection{\textbf{Centrally peaked cross-hair measurement setting}}

The crosshair measurement setting is given by the following:
{\small\begin{align}
p_1=\frac{1}{2}(1+\mathrm{tr}(\rho\sigma_x))=\frac{1}{2}(1+x), ~~~~~p_2=\frac{1}{2}(1-\mathrm{tr}(\rho\sigma_x))=\frac{1}{2}(1-x)\\
p_3=\frac{1}{2}(1+\mathrm{tr}(\rho\sigma_z))=\frac{1}{2}(1+z), ~~~~~p_4=\frac{1}{2}(1-\mathrm{tr}(\rho\sigma_z))=\frac{1}{2}(1-z)
\end{align}}
Therefore the posterior density from which we wish to sample from is given by 
$
(p_1p_2p_3p_4)^{10}=(\frac{1}{16}(1-x^2)(1-z^2))^{10}.
$

\begin{figure}[h]
\centering
\subfigure{
\label{fig:first}
\includegraphics[scale=0.435]{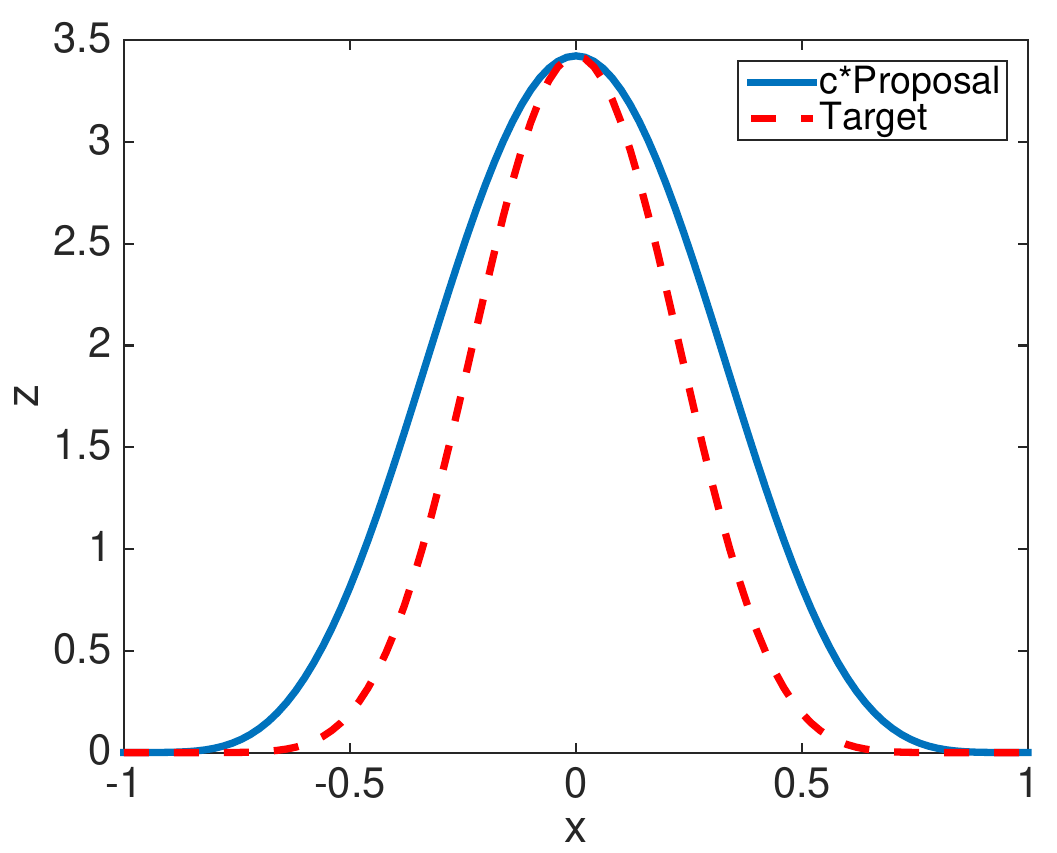}}
\qquad
\subfigure{
\label{fig:second}
\includegraphics[scale=0.435]{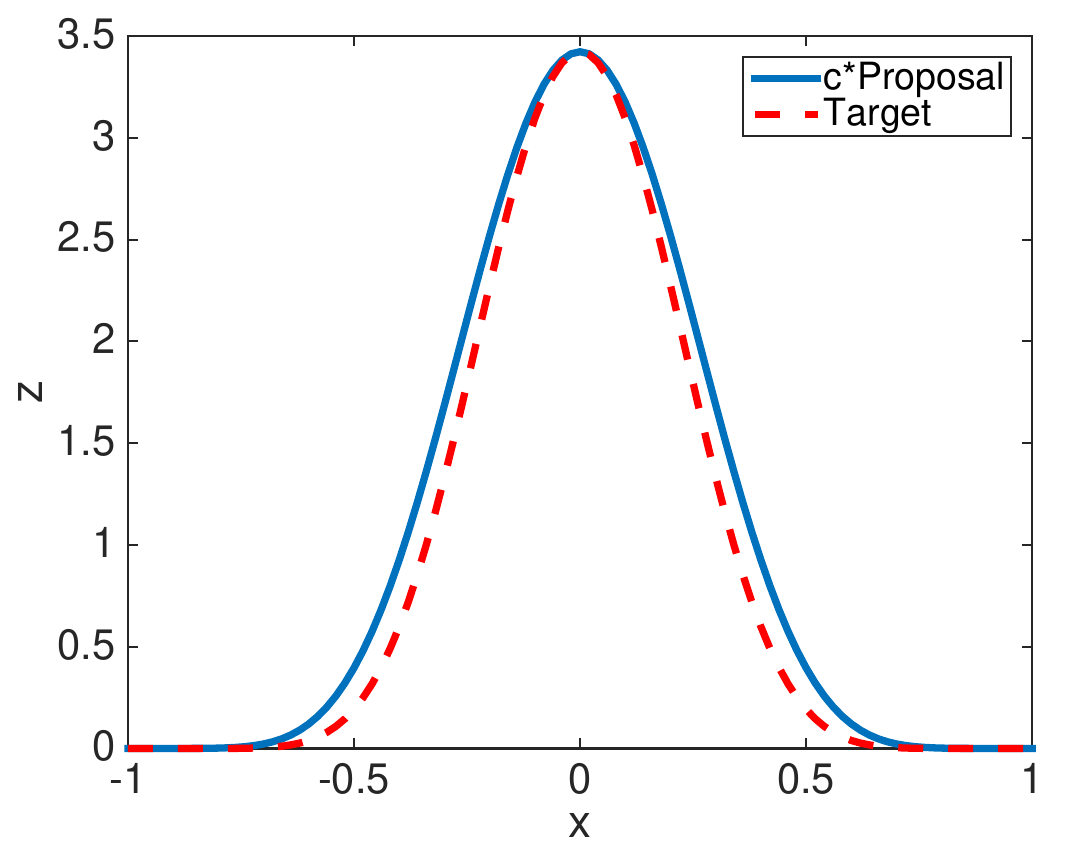}}
\caption{(a)This plot cross-section shows sub-optimal matching between proposal and target distribution for cross-hair measurement conjugate prior. Here number of columns is 13, the proportion of Uniform distribution is 0.0001.(b)This plot cross-section shows good matching between proposal and target distribution for cross-hair measurement conjugate prior. Here number of columns is 18, the proportion of Uniform distribution is 0.0001.}
\end{figure}

 With this simple example of a centred distribution, we show how the number of columns can be adjusted to make the proposal distribution closely match the target distribution to have good acceptance rates.  For performing the rejection sampling using our proposal distribution, we add a small amount of the uniform distribution to the Wishart like distribution to avoid the singularity while calculating the value of `C' . Next, after choosing the appropriate proposal distribution, we do the rejection sampling and obtain the target distribution. Adding only a very small proportion of uniform distribution such as $0.2$ percent to a Wishart distribution with $23$ number of columns gives the best acceptance rate of about $80$ percent for the example used here. This example shows that this algorithm is flexible to change according to any arbitrary height of the target distribution, therefore can be applied to any such scenario, with even much higher number of detector clicks. With the next example, we show that this holds true for the complex Bloch Ball. In the example number $2$, we sample according to the conjugate prior corresponding to the cross hair measurement in the Bloch Ball. The crosshair measurement setting is given by the following:
{\small\begin{align}\nonumber
p_1=\frac{1}{2}(1+\mathrm{tr}(\rho\sigma_x))=\frac{1}{2}(1+x), ~~~~~p_2=\frac{1}{2}(1-\mathrm{tr}(\rho\sigma_x))=\frac{1}{2}(1-x)\\ \nonumber
p_3=\frac{1}{2}(1+\mathrm{tr}(\rho\sigma_z))=\frac{1}{2}(1+y), ~~~~~p_4=\frac{1}{2}(1-\mathrm{tr}(\rho\sigma_z))=\frac{1}{2}(1-y)\\ \nonumber
p_5=\frac{1}{2}(1+\mathrm{tr}(\rho\sigma_z))=\frac{1}{2}(1+z), ~~~~~p_6=\frac{1}{2}(1-\mathrm{tr}(\rho\sigma_z))=\frac{1}{2}(1-z) \nonumber
\end{align}}
Therefore the posterior density from which we wish to sample from is given by 
$
(p_1p_2p_3p_4p_5p_6)^{10}=(\frac{1}{64}(1-x^2)(1-y^2)(1-z^2))^{10}.
$
In this case, we perform the sampling in exactly the same way as before, and compare the acceptance rates for different number of columns and proportions of uniform distributions. The algorithm perform well here, with the best acceptance rate of about $75$ percent for $11$ number of columns and $0.2$ percent of mixing uniform distribution. Until now we investigated the flexibility with respect to the height of the target distribution. In the next sections, we develop it for cases where we have flexibility with respect to both the position and height of the target distribution. 
\subsection{Trine measurement setting}
Here, we apply our algorithm for the trine measurement setting, where we have unequal number of detector clicks. The trine POM has three POM outcomes. These outcomes are constructed from three pure states symmetrically arranged in the real unit disc (X-Z plane) of the Bloch sphere, 
subtending angles of $2\pi$ between pairs of states. The probabilities of the outcomes corresponding to the trine measurement setting is given by the following:
$
p_1=\frac{1}{3}(1+r\mathrm{cos}\phi),
p_2=\frac{1}{3}(1+r\mathrm{cos}(\phi-\frac{2\pi}{3})),
p_3=\frac{1}{3}(1+r\mathrm{cos}(\phi-\frac{4\pi}{3}))
$.
Therefore the posterior density from which we wish to sample from is given by 
$
(p_1^7p_2^{10}p_3^{13})=\frac{1}{3^{30}}(1+r\mathrm{cos}\phi)^7(1+r\mathrm{cos}(\phi-\frac{2\pi}{3}))^{10}(1+r\mathrm{cos}(\phi-\frac{4\pi}{3}))^{13}.
$
\begin{figure}[h]
\centering
\subfigure{
\label{fig:first}
\includegraphics[scale=0.45]{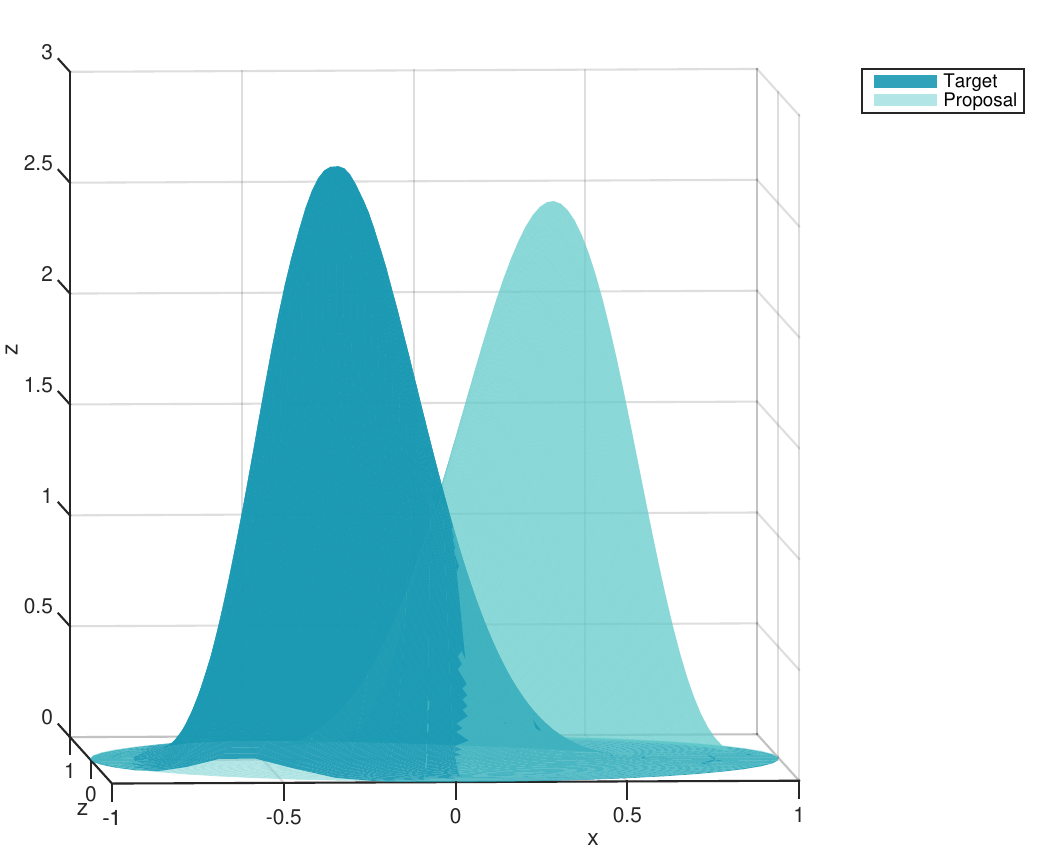}}
\qquad
\subfigure{
\label{fig:second}
\includegraphics[scale=0.45]{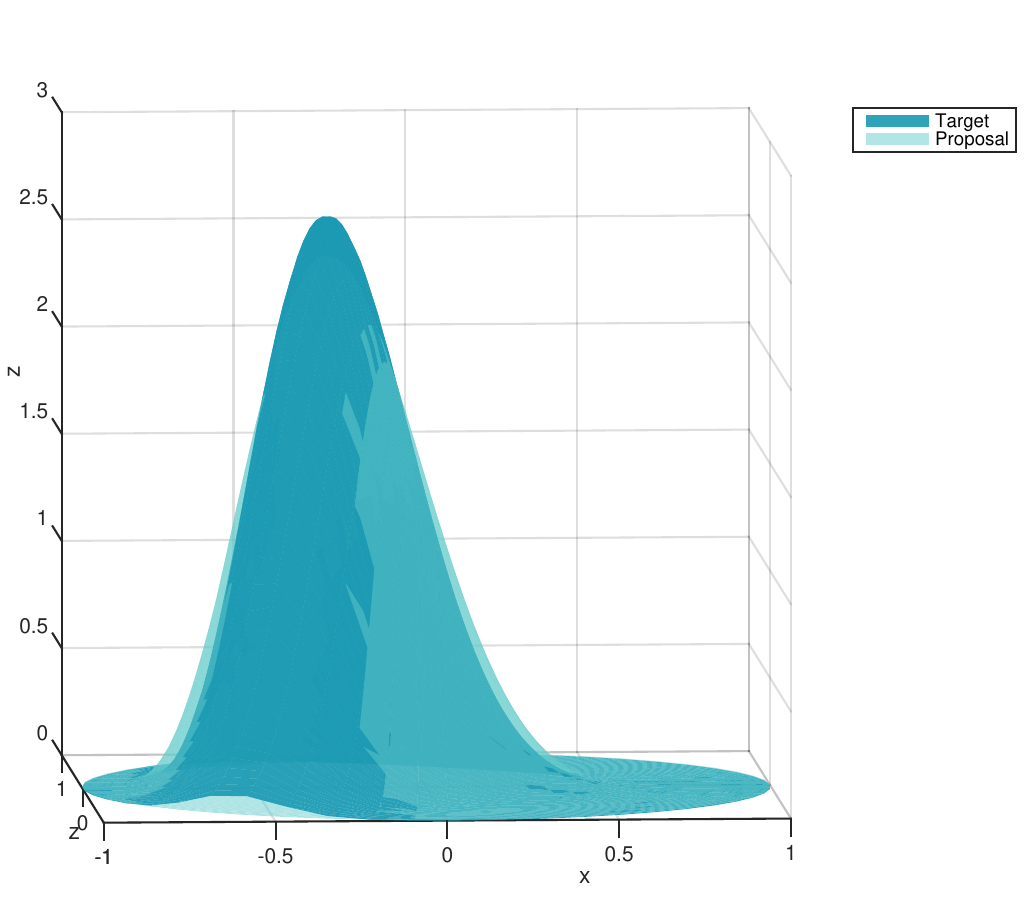}}
\caption{Plots of the matching of the target distribution (dark blue) with the proposal distribution (light blue) after an appropriate rotation. The right figure shows the achieved close match. }
\end{figure}
\begin{figure}[h]
\centering
\includegraphics[scale=1]{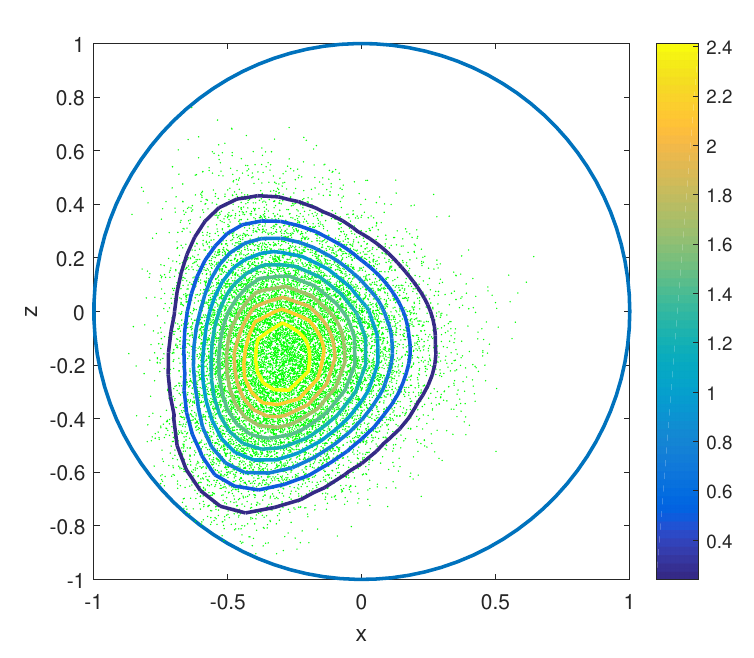}
\caption{Fig.(a) Plot of sampled points of the trine target distribution. }
\end{figure}
\begin{figure}[h]
\centering
\includegraphics[scale=0.95]{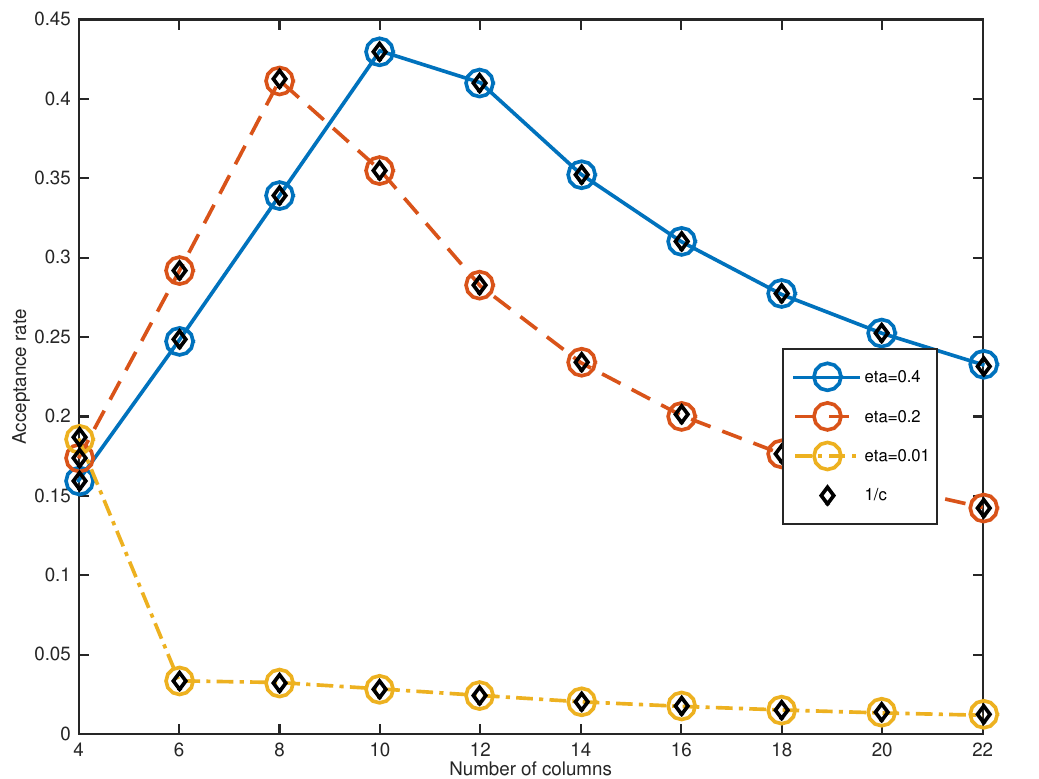}
\caption{Fig.(a) Plots of the acceptance rates of sample points vs the number of columns for different proportions of uniform distribution added to the proposal distribution. }
\end{figure}
With unequal number of detector clicks, the peak of the target distribution is not at the center. Therefore, to obtain proposal sample such that we have good acceptance rate 
by matching the target distribution, we take a nonzero mean value. This mean value can be fixed for different number of columns as has been described in the previous section, such that we match the peak of the target. With this fixing, we obtain different acceptance rates for different values of number of columns, as has been shown in the plot. From there we see that it still has the flexibility
and the best acceptance rate is very close to $50$ percent when the number of columns is $10$ and proportion of uniform distribution is about $0.4$. This acceptance rate is quite 
large compared to that when one uses only uniform distribution and has an acceptance rate of less than $5$ percentage. 
\begin{center}
\begin{table}
    \begin{tabular}{|  p{2.5cm} |  p{2.5cm} |  p{2.5cm} | p{2.5cm}  |p{2.5cm} | p{2.5cm}  |}
    \hline
    N & 4 & 6 & 8 & 10 & 12\\ 
    \hline
    $\mu$ & 0.326882 & 0.481645 & 0.548839 & 0.587081 & 0.611833\\ 
    \hline
    N & 14 & 16 & 18 & 20 & 22\\ 
    \hline
    $\mu$ & 0.629177  & 0.642010  & 0.651891 & 0.659734 & 0.666111\\
    \hline
\end{tabular}
\caption{This table shows the values of the mean corresponding to the different number of columns to fix the peak position at a desired position.}
\end{table}
\end{center}

\subsection{\textbf{Tetrahedron measurement setting}}

The tetrahedron measurement setting is an informationally complete four outcome POM. The outcome probabilities are given by the following : 
{\small\begin{align}\nonumber
p_1=\frac{1}{4}(1+\frac{1}{\sqrt{3}}(x+y+z))~~~~~~~p_2=\frac{1}{4}(1+\frac{1}{\sqrt{3}}(x-y-z))\\ \nonumber
~~p_3=\frac{1}{4}(1+\frac{1}{\sqrt{3}}(-x+y-z))~~~~~p_4=\frac{1}{4}(1+\frac{1}{\sqrt{3}}(-x-y+z))\nonumber
\end{align}}%
where, $\mathrm{x}=\mathrm{tr}(\sigma_x\rho)$, $\mathrm{y}=\mathrm{tr}(\sigma_y\rho)$ and $\mathrm{z}=\mathrm{tr}(\sigma_z\rho)$.
Posterior distribution is defined as $p_1^{k_1}p_2^{k_2}p_3^{k_3}p_4^{k_4}$ as before. Some examples are discussed in the following sections.

\begin{figure}[h]
\centering
\includegraphics[scale=0.85]{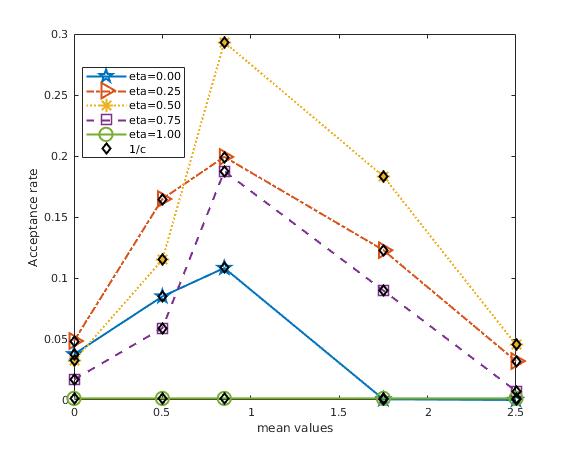}
\caption{Fig.(a) Plot of acceptance rates for different values of mean and number of columns.
The data is for the 3 dimensional tetrahedral measurement according to detector click sequence $\{12,7,21,10\}$ as shown in the example. }
\end{figure}
\begin{figure}[h]
\centering
\includegraphics[scale=0.5]{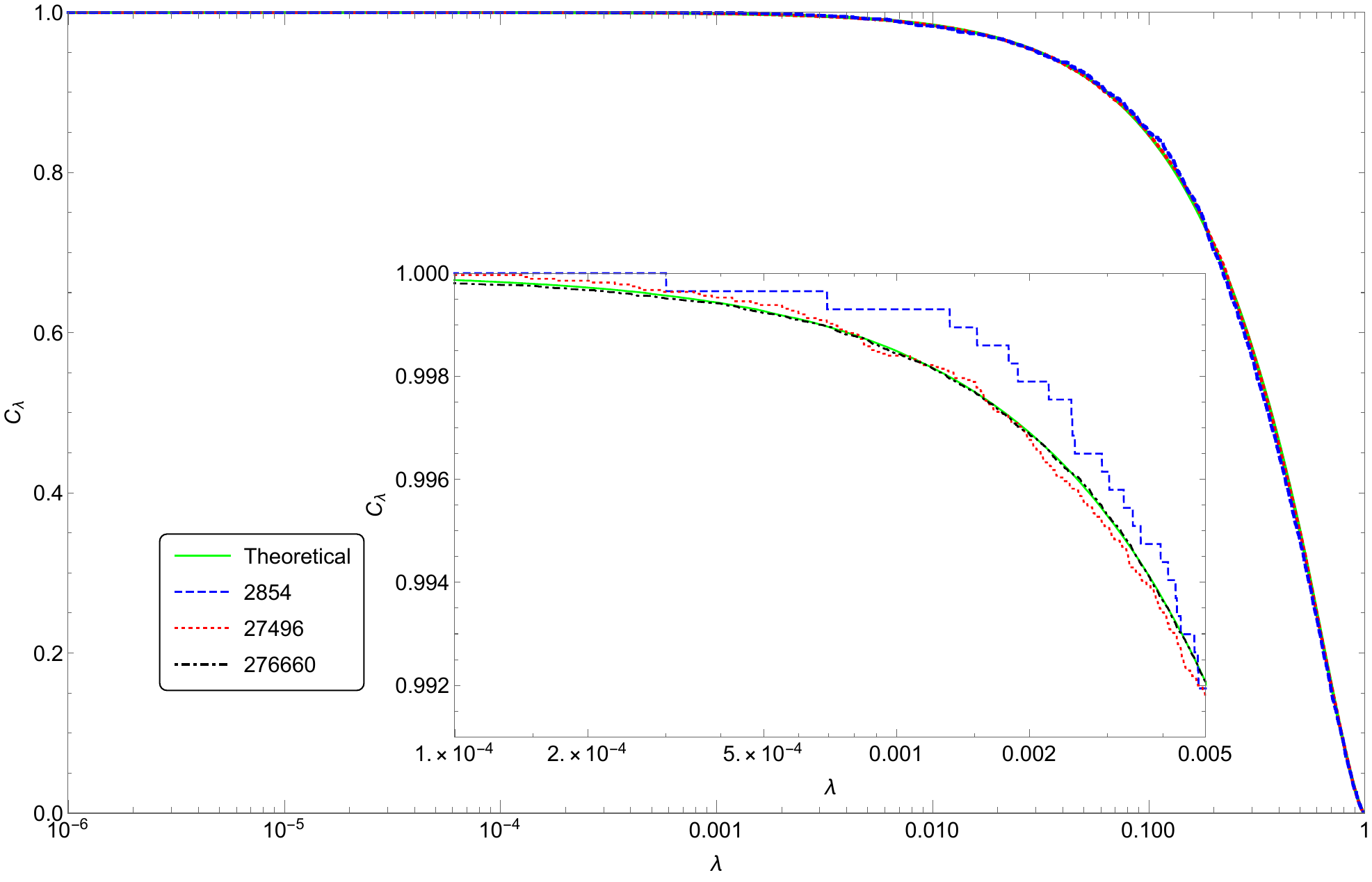}
\caption{Fig.(a) Plot of credibility for different sample sizes vs the theoretical credibility calculated from the size of the bounded likelihood region. 
The data is for the 3 dimensional tetrahedral measurement setting according to detector click $\{5,20,23,7\}$. The different colours are for different sample 
sizes as shown by the legends. 
}
\end{figure}

\subsubsection{\textbf{Moderate value of conjugate prior on surface of Bloch Ball}}

The first example we discuss pertains to the detector click sequence $\{12,7,21,10\}$, with total number of detector clicks $50$. The peak of this distribution is at $r=0.72332, \theta=2.18302, \phi=1.92956$. 
We note that in this case, since one has a significant amount of nonzero target distribution on the boundary, therefore, the Wishart distribution which is zero on the boundary, i.e., for number of columns 
greater than $2$ performs poorly in comparison to earlier examples. Also, for similar reason sampling from the uniform distribution performs even poorer than that, rendering these 
methods inflexible. Indeed, we can have acceptance rate only about $5$ percent with these methods and even lower with that of uniform distribution. Therefore, in contrast to the previous methods, here we mix a distribution with number of columns $2$, yet a nonzero mean value, such that we can have a nonzero value of the distribution on the boundary. with a distribution which matches the peak position inside the Bloch ball. Indeed this improves the acceptance rate to a very good value of about $30$ percent. The parameters used in this proposal distribution are given in the plot of the acceptance rate. We choose $4$ number of columns in this example since this gives us good acceptance rates. We see that when we mix equal proportions of these two distributions, we get good acceptance rates over a wide range of the values of the mean. For the Wishart which is zero on the boundary we choose the mean, such that it coincides with that of the conjugate prior after appropriate rotation. The best 
acceptance rate of about $30$ percent occurs when we have mixed equal proportions and chosen the mean value of the Wishart non-zero on the boundary to be 0.85. 
\subsubsection{Very high value of the conjugate prior on surface of Bloch Ball}
\begin{center}
\begin{table}[h]
    \begin{tabular}{|  p{2.5cm} | p{2.5cm}  |  p{2.5cm} |  p{2.5cm} | p{2.5cm}  |p{2.5cm} | }
     \hline
   $\mu$ & 0& 0.5 & 1.15 & 1.5 & 2.0 \\ 
    \hline
    AR\% & 1.6 & 5.65 & 20.52 & 10.68 & 0.2\\ 
    \hline
   $\mu$ & 0.5& 1.0 & 1.12 & 1.5 & 2.0 \\ 
    \hline
    AR\% & 17.85 & 30.26 & 32.09 & 20.18& 13.38\\  
    \hline
\end{tabular}
\caption{The upper table shows the acceptance rate percentage for different mean values when we choose to sample using only one Wishart which is non-zero on the boundary.
This pertains to the tetrahedral measurement setting with detector click sequence $\{5,20,23,7\}$.The lower table shows the acceptance rate percentage for different mean values (of Wishart with number of columns as $3$)  when we choose to sample using 0.39 proportion of Wishart which is zero on the boundary (number of columns $3$ and the different mean values as shown in the table )
with 0.61 proportion of Wishart which is zero on the boundary with mean value of 1.2.
This pertains to the tetrahedral measurement setting with detector click sequence $\{5,20,23,7\}$. Among other combinations of mean values and proportions, this
is the most optimal.}
\end{table}
\end{center}
In this example we discuss the detector click sequence $\{5,20,23,7\}$ with total detector clicks being $55$. The peak of this distribution is at $r=0.989365, \theta=1.73063, \phi=3.10935$. 
In this case, if we use only the Wishart distribution which is non-zero on the boundary, then we can adjust the appropriate proposal distribution by just fixing the value of the mean which then fixes the height of the peak on the Bloch Ball.  After fixing mean to fix the height of the proposal
distribution to match that of the target distribution, we just need to unitarily transform or rotate to match the peak position of the conjugate prior on the boundary of the Bloch Ball. Using this method, one improves the acceptance rate to a very reasonable value of over $21$ percent. The values of the acceptance rate for the different values of mean are given in Table 5. In a second method, if one uses a mixture of Wishart which is zero on the boundary and Wishart which is non-zero on the boundary, one can find a proposal distribution by adjusting the number of columns to $3$ to improve the acceptance rate to over $32$ percent. From the plot it is clear that we get the best acceptance rate when we mix $0.4$ proportion of Wishart with number of columns $3$ and $0.6$ proportion of Wishart with number of columns $2$ with both mean values fixed at $1.18$. This shows that the algorithm is rather flexible to suit various requirements. The uniform distribution still fares very poorly in this respect with an acceptance rate of about 3 percent.

\subsubsection{\textbf{Peaks of the conjugate prior on surface of Bloch Ball}}

In this section, we discuss two examples where the $\rho_{MLE}$ is purely unphysical and actually lies outside the boundary of the state space. The first example what we discuss pertains to the detector click sequence $\{5,20,33,7\}$ with total detector clicks being $65$. 
The peak of this distribution is at $r=1.0, \theta=1.88154, \phi=2.94285$. 
In this example, we can have acceptance rate of only about $0.65$ percent with uniform distribution alone. On the other hand, we know that fixing the value of the mean for the Wishart distribution with number of columns $2$
changes the height of the peak on the Bloch Ball. Using this distribution, one improves the acceptance rate to a very reasonable value of over $21$ percent. We give some examples of the acceptance rates with respect to chosen height and the value of the mean in the table. The second example we show that even if we make this kind of distribution more peaked, even then there is no scaling of the 
acceptance rate . For this we take the example of having detector clicks $\{55,50,13,10\}$ with total detector clicks being $128$. 
The peak of this distribution is at $r=1.0, \theta=0.096255, \phi=1.49475$. 
We can have acceptance rate only about $0.2$ percent with 
uniform distribution alone, however using our algorithm, one improves the acceptance rate to a very reasonable value of over $21$ percent again. Specifically in this case it 
is hundred times more efficient. 
\begin{center}
\begin{table}[h]
    \begin{tabular}{|  p{2.0cm} | p{2.0cm}  |  p{2.0cm} |  p{2.0cm} | p{2.0cm}  |p{2.0cm} | p{2.0cm} |}
     \hline
   $\mu$ & 0& 0.5 & 1.0 & 1.5 & 2.0 & 2.5\\ 
    \hline
    AR\% & 0.6 & 2.22 & 10.75 & 21.20 & 2.74 & 0.1 \\ 
        \hline
   $\mu$ & 0& 0.5 & 1.0 & 1.8 & 2.0 & 2.5\\ 
    \hline
    AR\% & 0.2 & 0.8 & 9.32 & 21.01 & 2.74 & 0.5 \\ 
    \hline
\end{tabular}
\caption{The upper table shows the acceptance rate percentage for different mean values when we choose to sample using only one Wishart which is non-zero on the boundary.
This pertains to the tetrahedral measurement setting with detector click sequence $\{5,20,33,7\}$. The lower table shows the acceptance rate percentage for different mean values when we choose to sample  using only one Wishart which is non-zero on the boundary.
This pertains to the tetrahedral measurement setting with detector click sequence $\{55,50,13,10\}$.}
\end{table}
\end{center}
\section{Performance with respect to CPU time}
In this section we break down the time required for the performance of the sampling algorithm. For this, we note that there are three important steps in performance of 
the algorithm. The first step is finding the peak and peak height of the target distribution. On a regular desktop and CPU, this takes about a few
seconds, since this is only for qubits. The second important step is finding an appropriate proposal distribution to match the target distribution. This involves fixing the peak position and the peak height of the proposal distribution. It is better to first match the peak height close to that of the target distribution. This takes
about a few seconds together with then fixing the peak position. Also note that all these steps can be automatized very easily in a single program. The next step then 
is to perform the acceptance rejection sampling. The last steps takes about
40 seconds, to generate a million sample points. Therefore in total we can obtain a high quality random sample according to any target distribution in less than a minute for a million of sample points. Upon comparing with the time taken in Monte Carlo algorithms, the typical time taken for generating about a million sample points corresponding to an off-centered conjugate prior of a crosshair measurement setting in the real plane takes about 8 hours whereas to generate a million point according to that setting in the Bloch Ball takes almost double of that which is about fifteen to sixteen hours. Therefore, our algorithm fairs significantly better than the time taken in Monte Carlo algorithms. Next, we compare that with uniform distribution. For this let us consider the examples here. It is straightforward to notice that the time taken to sample the points is directly proportional to the acceptance rates. Therefore for the cases where the conjugate prior is quite sharply peaked compared with a flat distribution, our algorithm should perform 
progressively better in time than that for the uniform distribution. 
It is easy to see that this effect will 
get more pronounced when one has more experimental data and the conjugate prior becomes more peaked.
\begin{center}
\begin{table}[h]
    \begin{tabular}{|  p{3.75 cm} | p{3.75cm}  |  p{3.75cm} |  p{3.75cm} |}
     \hline
   $Method$ & MCMC & HMC&Wishart \\ 
        \hline
    Time & 15 hours & 15 hours&5 min \\ 
     \hline
    IID & No & No &Yes \\ 
    \hline
\end{tabular}
\caption{The table shows the time taken for generation of a sample of million random density matrices according to e tetrahedral measurement setting with total detector clicks of 50, by different methods: MCMC stands for Markov Chain Monte Carlo, HMC stands for Hamiltonian Monte Carlo, IID stands for independent and identically distributed.}
\end{table}
\end{center}

\section{\textbf{Sampling for two qubits for some special cases}}

The specific algorithm used in the examples here uses the unitary transformation or rotation of sample points to match the peak of the target distribution and has been performed in qubit space. However, as we move on from the qubit space to that of the two qubits, the Bloch vector does not retain its nice properties. One cannot use rotation in higher dimension, since after rotation of Bloch vector in higher dimension we may not get a physically valid density matrix. One may use unitary mapping to still perform an algorithm with proposal distributions generated according to the Wishart distribution. The unitary operators when acting on the mean matrix generate a set of mean matrices, which then shifts the position of the peak around various positions in the state space of two qubits. However, this may not be the most optimal method in the sense that it does not cover the entire state space. We keep this for further research. Apart from this however,  the acceptance rejection sampling method is not particularly suited for sampling in much higher dimensions such as for four qubits, as is captured by the concept of ``curse of dimensionality", where the acceptance rate becomes increasingly low to an extent that makes this algorithm impractical. As a result, for such higher dimensions one has to resort to methods such as the Monte Carlo sampling or Gibbs sampling \cite{Han}. We also keep this direction for further refinement.

\section{Conclusions and future directions}

The work presented in this article is concerned with finding the closed form expression of probability distribution function for random quantum states derived using the non-central Wishart distribution and demonstrating its usefulness with a sampling algorithm useful in the context of experimental data for quantum state estimation. With this motivation in mind, we have found out the closed form expression for the distribution of random quantum states corresponding to non-central Wishart distribution with rank one non-zero mean matrix and general covariance matrix for the case of both real and complex Hilbert space. We name this the non-zero mean quantum Wishart distribution. This extends the previous works on random quantum states to newer territories. We know that the theory of random quantum states form a very important part of study in the field of quantum information science. As a result, the results derived here is expected to be of use in future in different areas of quantum information science where the random quantum states are important. Along this line of research, for future directions, the open question remains to find out the closed form expression for the distribution of random quantum states corresponding to non-zero higher rank mean matrix.

Aa an application, we have proposed an algorithm to sample random quantum states according to a target distribution which may have an arbitrary shape in the probability space but well behaved, best suited for single peaks. Similar work has been presented in \cite{Han}, however our newfound distribution function here offers a lot more flexibility. This algorithm is efficient with respect to some previously existing algorithms. For example, in contrast to the usage of uniform distribution which is not flexible to suit a variety of situations for example for the cases of very sharp peaks, we have used some parameters pertaining to the non zero mean quantum Wishart distribution that lets us tweak the algorithm to adapt to an arbitrary landscape of target distribution function in the qubit space. Also, in contrast to the Monte Carlo algorithm, our algorithm generates sample points that do not have any correlation between them, like that proposed in \cite {Shang,Seah}. Not only this, it is also very CPU time cheap. We emphasize that the the lowest acceptance rate that we got using these distributions and their combinations in various scenarios is $20$ percent and has very nice scaling properties. However, for distributions with peaks situated inside and having non-zero value on the boundary,
it is evident that one can improve and optimize the performance by choosing proper parameters in a systematic way. Therefore, this algorithm offers advantages over the existing ones. Not only this, for target distributions with peaks on the boundary or some finite non-zero value on the boundary, we can use various parameters in the non-zero mean quantum Wishart distribution to get good acceptance rates of high quality samples of random quantum states. Moreover, each of steps in the algorithm can be automated to suit various scenarios which shall further reduce the time and any amount of complexity of performing the algorithm, for example in the form of an open source program in Mathematica, Matlab or C++ etc. As a result, we keep the task of writing out an open source program with such automated features as a future direction. On another note, we note that it does not work very well for higher dimensions, where newer tools may be required to adapt this algorithm there. We keep this direction for further research, see also \cite{Han}. Also, it will be an interesting direction to see if the reference distributions can be used in conjunction with Monte Carlo sampling or Gibbs sampling methods for sampling in higher dimensions. Another good potential direction in future is the application of the non-zero mean quantum Wishart distribution is the application in the multiple input and multiple output quantum channels in the way that the normal non-zero mean Wishart distribution has found use in the multiple and multiple output classical channels in the classical communication theory in information theory. This seems to be a promising direction of future research. As a result, the work presented here adds substantial value to the quantum information science community and we hope that it will be useful to the researchers in this particular field.

\section*{Acknowledgments}

SB acknowledges funding from Korea Institute of Science and Technology. S. B. acknowledges support from the National Research Foundation of Korea (2020M3E4A1079939, 2022M3K4A1094774) and the KIST institutional program (2E31531). SB acknowledges discussions with B. Englert, H. Rui, L. Weijun and N.G.H Khoon while her postdoctoral tenure at CQT (NUS), Singapore where a part of this work was done.

\end{document}